\documentclass[fleqn,usenatbib,useAMS]{mnras}
\usepackage{graphicx}   
\usepackage{amsmath}    
\usepackage{amssymb}    
\usepackage{multicol}        
\usepackage{bm}         
\usepackage{pdflscape}  
\usepackage[T1]{fontenc}
\usepackage{ae,aecompl}
\usepackage{newtxtext,newtxmath}

\def\aap{A\&A}
\def\apjl{ApJ}
\def\apj{ApJ}
\def\apjs{ApJS}

\def\mnras{MNRAS}
\def\nat{Nature}
\def\pasp{PASP}

\title[Ultramassive White Dwarfs]{The Merger Fraction of Ultramassive White Dwarfs }
\author[Kilic et al.]
{Mukremin Kilic$^1$, Adam G. Moss$^1$, Alekzander Kosakowski$^2$, P. Bergeron$^3$, 
\newauthor  Annamarie A. Conly$^4$, Warren R. Brown$^5$, Silvia Toonen$^6$, Kurtis A. Williams$^7$, P. Dufour$^3$\\
$^1$Homer L. Dodge Department of Physics and Astronomy, University of Oklahoma, 440 W. Brooks St., Norman, OK 73019, USA\\
$^2$Department of Physics and Astronomy, Texas Tech University, Lubbock, TX 79409, USA\\
$^3$D\'epartement de Physique, Universit\'e de Montr\'eal, C.P. 6128, Succ. Centre-Ville, Montr\'eal, QC H3C 3J7, Canada\\
$^4$Department of Astronomy, New Mexico State University, Las Cruces, NM 88003, USA\\
$^5$Smithsonian Astrophysical Observatory, 60 Garden Street, Cambridge, MA 02138, USA\\
$^6$Anton Pannekoek Institute for Astronomy, University of Amsterdam, 1090 GE Amsterdam, The Netherlands\\
$^7$Department of Physics \& Astronomy, Texas A\&M University-Commerce, Commerce, TX 75429, USA\\
}

\date{\ \ Submitted \today \vspace{-0.5cm}}
\pubyear{2022}

\begin{document}
\label{firstpage}
\pagerange{\pageref{firstpage}--\pageref{lastpage}}
\maketitle

\begin{abstract}

We search for merger products among the 25 most massive white dwarfs in the Montreal White Dwarf Database 100 pc sample
through follow-up spectroscopy and high-cadence photometry. We find an unusually high fraction, 40\%, of magnetic white
dwarfs among this population. In addition, we identify four outliers in transverse velocity and detect rapid rotation in five objects. Our results
show that $56^{+9}_{-10}$\% of the  $M\approx1.3~M_{\odot}$ ultramassive white dwarfs form through mergers. This fraction is
significantly higher than expected from the default binary population synthesis calculations using the $\alpha$-prescription (with
$\alpha \lambda = 2$), and provides further support for efficient orbital shrinkage, such as with low values of the common envelope efficiency.

\end{abstract}


\section{Introduction}

Binary stars are common, but binary white dwarfs are less so. The multiplicity fraction of A type stars, the dominant progenitor systems for white dwarfs
in the solar neighborhood, is around 45\% \citep{derosa14,moe17}, whereas that of the white dwarfs in the local 20-25 pc sample is significantly
lower at $\approx25$\% \citep{holberg16,hollands18}. This discrepancy is not simply due to observational biases against detecting a dim white dwarf companion close to a bright star, but instead it can be explained if  a significant fraction of the binary systems disappear
on or after the main-sequence phase through mergers, and form single stars that evolve into single white dwarfs \citep{toonen17,temmink20}. 

Accreting CO core white dwarfs as well as mergers of double white dwarfs can reach the Chandrasekhar limit and explode
as type Ia supernovae \citep{webbink84,iben84}. Sub-Chandrasekhar-mass white dwarfs can also detonate \citep[e.g..][]{shen18}. 
However, due to the steepness of the initial mass function, most merger events involve lower mass white dwarfs, and they do not lead
to explosive transient events. Instead they form single, more massive white dwarfs \citep{garciaberro12,schwab21}. Based on binary population
synthesis calculations, \citet{temmink20} estimated that between
about 10 to 30\% of all single white dwarfs are formed through binary mergers, with the majority of them involving the descendants of mergers between
post-main-sequence and main-sequence stars. However, the predicted merger fraction goes up to 30 to 45\% for 
all observable single white dwarfs with $M>0.9~M_{\odot}$ and within 100 pc of the Sun. 

Observational constraints on the merger fraction of single white dwarfs are scarce. \citet{maoz12}, \citet{maoz17}, and \citet{maoz18} used
a statistical method for characterizing the binary white dwarf population in the Sloan Digital Sky Survey and the ESO-VLT Supernova-Ia to
estimate that 8.5 to 11\% of all white dwarfs ever formed have merged with another white
dwarf. This is significantly higher than predicted from the binary population synthesis models, $\sim$1-3\%, of \citet{temmink20}, and if
true, has implications for the type Ia supernova progenitors \citep{maoz18,cheng20} and the gravitational wave foreground from the Galactic double white dwarf population
in the milli-Hertz frequency band \citep{korol22}.

\citet{kilic21} presented an analysis of the 25 most massive ($M \geq 1.3~M_{\odot}$) white dwarf candidates in the
Montreal White Dwarf Database 100 pc sample and concluded that at least 32\% of these white dwarfs are likely double white
dwarf merger products based on their kinematics, magnetism, or rapid rotation. However, only 10 of these objects currently have
spectral classification available, and interestingly, four are magnetic. 
\citet{tout08} and \citet{briggs15} argued that all strongly magnetic white dwarfs have a binary origin, as a
magnetic dynamo can be generated during a merger event though differential rotation within a common-envelope or an accretion disk.
In addition, only one of these targets, J1832+0856, had follow-up high cadence photometry available, which revealed a spin period
of only 353 s \citep{pshirkov20}. The fraction of merger products among the ultramassive white dwarfs may be much higher. 

Recently, \citet{caiazzo21} found a rotation period of 6.94 min in another object in this sample, J1901+1458, and
\citet{kilic21b} discovered photometric variations in J2211+1136 with a period of only 70 s, making it the fastest spinning
isolated white dwarf currently known. These rotation rates are consistent with the
predicted rates for single white dwarfs that formed from double white dwarf mergers \citep{schwab21}. 

Here we present the results from a spectroscopic and photometric follow-up survey of all 25 ultramassive white dwarfs identified
by \citet{kilic21}. We use the optical spectroscopy of the remaining targets to search for evidence of magnetism and unusual
atmospheric composition that would indicate a merger origin, and high-speed photometry to search for evidence of fast rotation.
We use these observations to obtain, for the first time, a reliable merger rate estimate for ultramassive white dwarfs.
We present the details of our follow-up observations in Section 2, model atmosphere analysis in Section 3, and the constraints
on photometric variability in Section 4. We discuss the merger fraction of ultramassive white dwarfs in Section 5, along with
the implications for the binary population synthesis models, and conclude in Section 6.

\section{Observations}

\subsection{Spectroscopy}

We obtained follow-up optical spectroscopy of 13 targets using the Gemini North and South 8m telescopes equipped
with the Gemini Multi-Object Spectrograph (GMOS) as part of the queue programs GN-2022A-Q-303 and GS-2022A-Q-106.
We used the B600 grating and a $1\arcsec$ slit, providing wavelength coverage from 3670 \AA\ to 6800 \AA\
and a resolution of 2 \AA\ per pixel in the $4\times4$ binned mode. 

Two additional targets, J0329-2123 and J0426-5025, could not be observed at Gemini during the 2022A semester. We obtained
spectra for these two targets at the 6.5m Magellan telescope with the MagE spectrograph. We used the $0.85\arcsec$ slit, providing
wavelength coverage from about 3400 \AA\ to 9400 \AA\ with a resolving power of R = 4800. To extend the wavelength coverage for
one of our Gemini targets with an intriguing spectrum, J1819$-$1208, we obtained additional MagE observations that confirmed
the hot DQ spectral classification.

\subsection{High-cadence Photometry}

We acquired high speed photometry of 15 of our targets between 2021 October and 2022 July
using the APO 3.5m telescope with the Agile frame transfer camera \citep{mukadam11} and the BG40 filter.
We obtained back-to-back exposures of 10-30 s over 2 hours for most of the objects, but we were limited to
$\sim$1-1.5 hours of observation time for three targets, J1116$-$1603, J1140+2322, and J1329+2549. We binned the CCD by $2\times2$,
which resulted in a plate scale of $0.258\arcsec$ pixel$^{-1}$. 

We obtained simultaneous $g-$ and $i-$band observations of four targets using the dual-channel frame-transfer camera Zorro on the Gemini
South telescope as part of the program GS-2022A-Q-303. We obtained 10 s long back-to-back exposures of each target over an hour.
Zorro provides imaging over a $60\arcsec$ field of view in the wide-field mode with a plate scale of $0.07\arcsec$ pixel$^{-1}$.

We obtained additional time-series observations of two targets with the ProEm frame-transfer CCD on the McDonald Observatory 2.1m
Otto Struve Telescope at Cassegrain focus. We used the BG40 filter with 15-20 s exposures. 

Including the APO observations of the three magnetic white dwarfs from \citet{kilic21b}, we have so far observed
20 of the 25 white dwarfs in our sample with the frame transfer cameras at the APO, Gemini, and McDonald Observatory telescopes.
Two additional targets have high-cadence photometry published in the literature \citep{pshirkov20,caiazzo21}. Hence,
only three of our targets, J0049$-$2525, J0426$-$5025, and J1727+3831, currently lack follow-up high-speed photometry.

\section{Model Atmosphere Analysis}

\subsection{The Fitting Method}

\citet{kilic21} provided model atmosphere fits to all 25 ultramassive white dwarfs with $M\geq1.3~M_{\odot}$ in the
Montreal White Dwarf Database. However, they had spectroscopy available for only 10 of these targets. Here we revisit
the model atmosphere analysis of the 15 targets with recently obtained Gemini and Magellan follow-up spectroscopy.

We use the photometric technique, and use the SDSS $u$ and Pan-STARRS $grizy$
photometry along with the Gaia EDR3 parallaxes to constrain the effective temperature and the solid angle, $\pi (R/D)^2$,
where $R$ is the radius of the star and $D$ is its distance. Given precise distance measurements from Gaia, we
constrain the radius of each star directly, and therefore its mass based on the evolutionary models for a given
core composition. The details of our fitting method, including the model grids used are further discussed in \citet{bergeron19},
\citet{genest19}, \citet{blouin19}, and \citet{kilic20,kilic21}. 

\subsection{DA White Dwarfs}

\begin{figure*}
\centering
\includegraphics[width=2.05in]{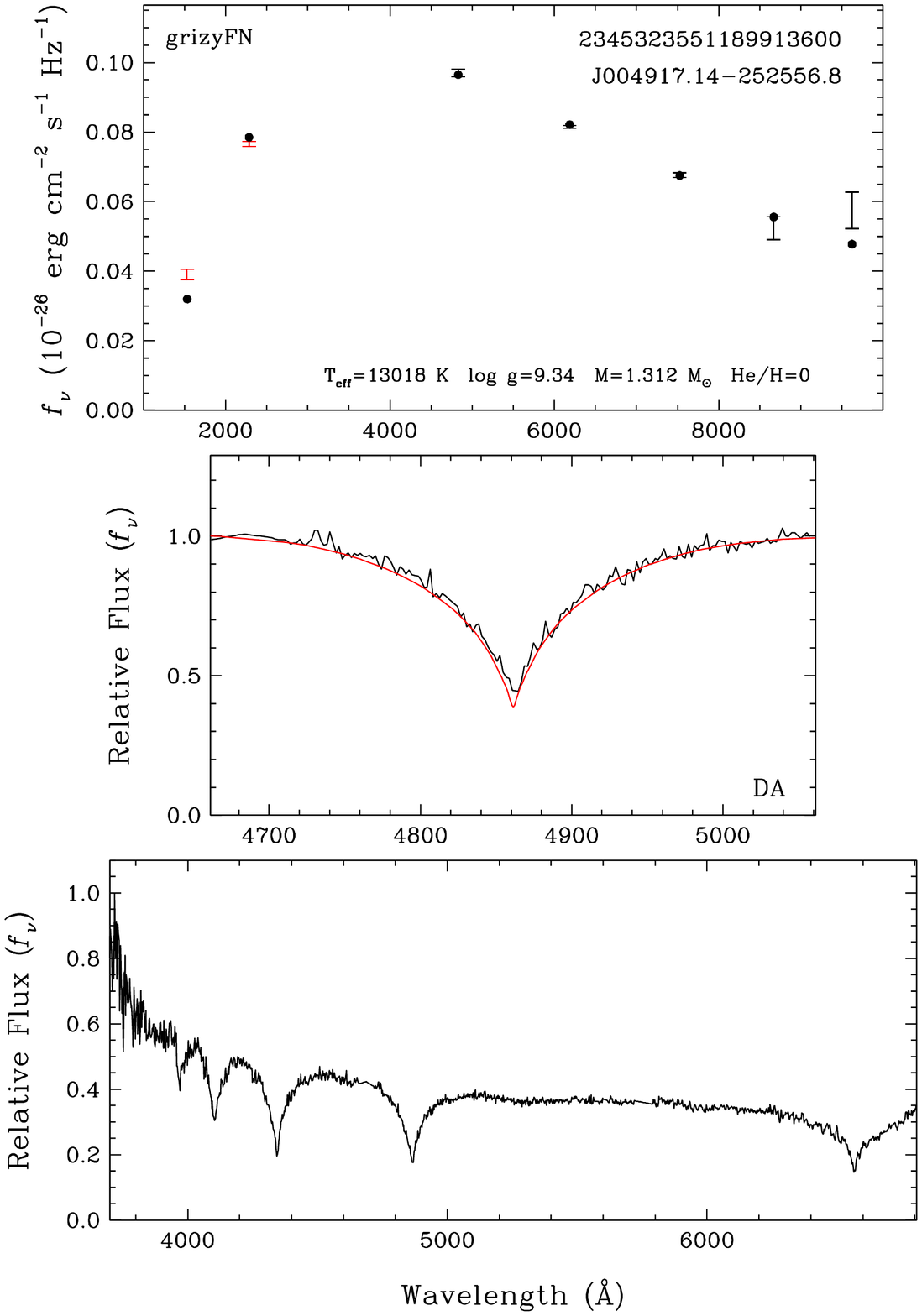}
\includegraphics[width=2.05in]{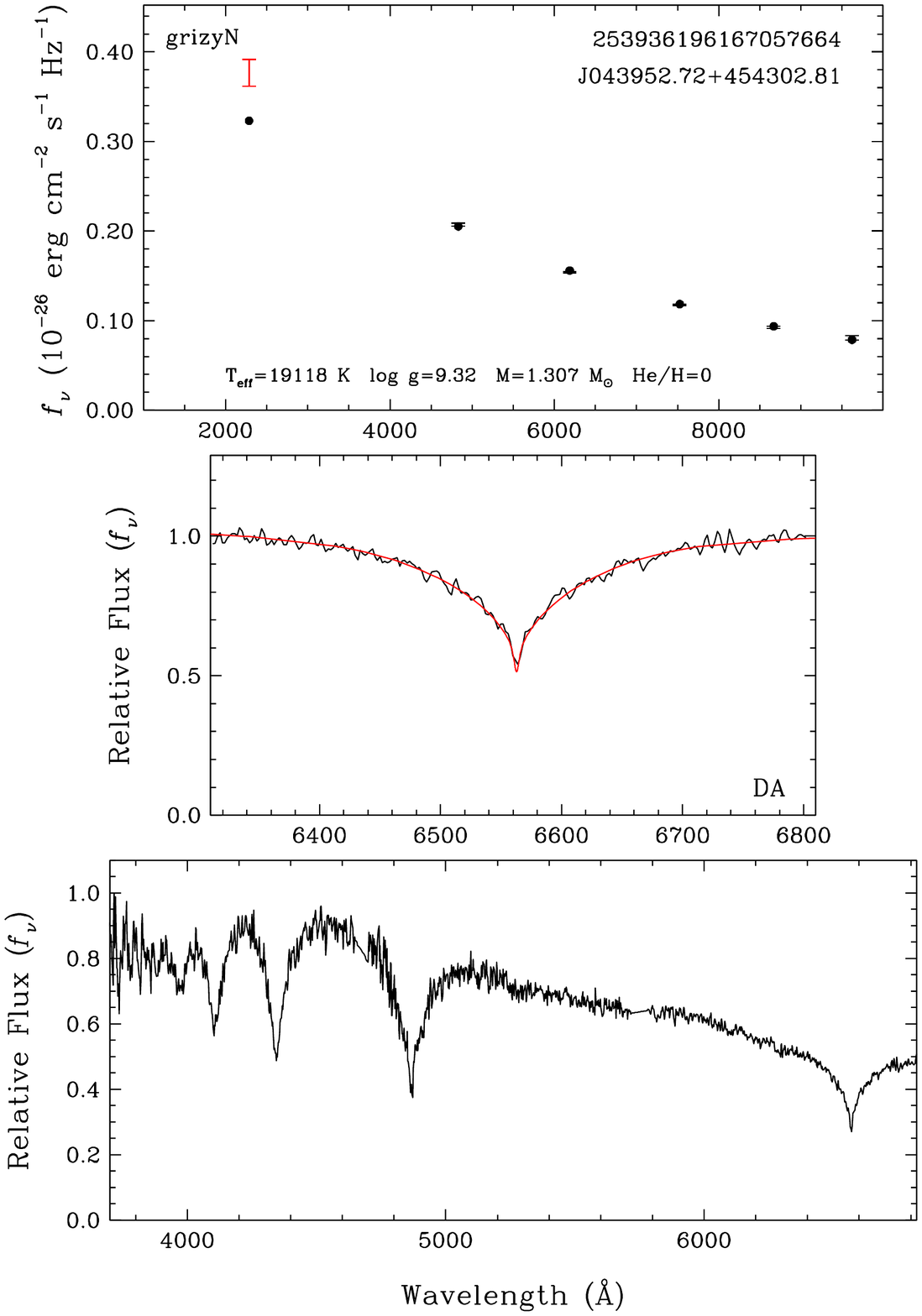}
\includegraphics[width=2.05in]{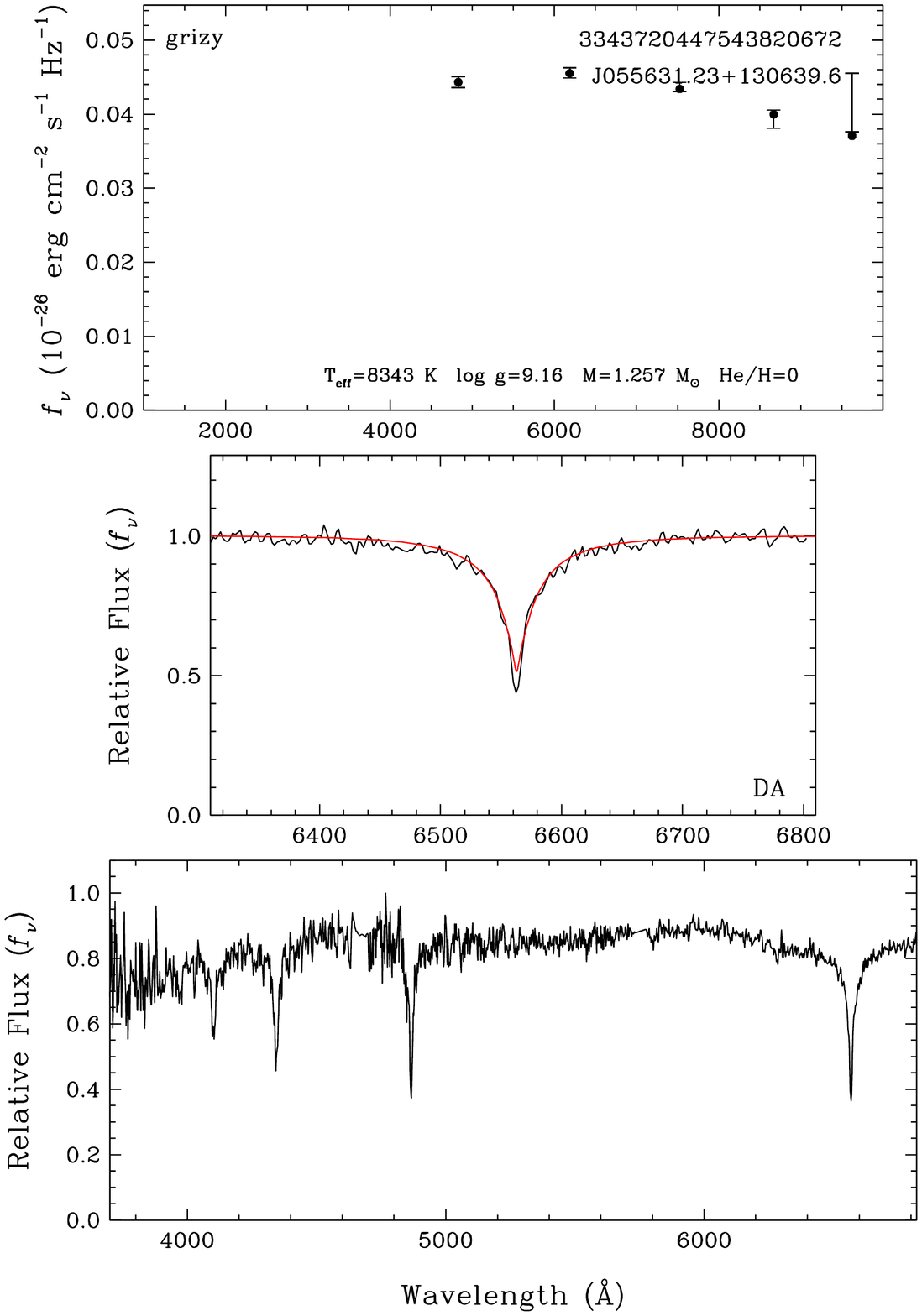}
\includegraphics[width=2.05in]{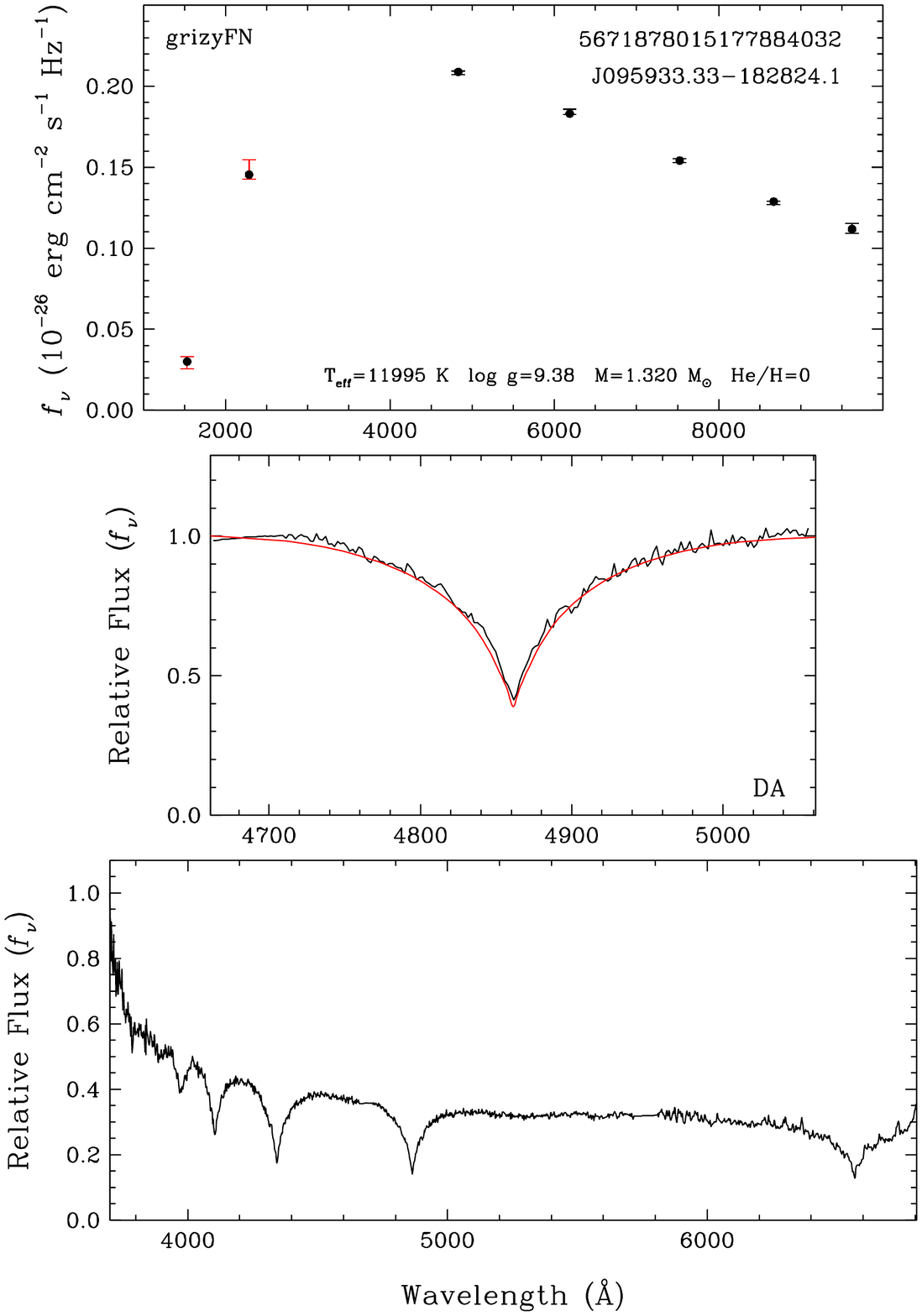}
\includegraphics[width=2.05in]{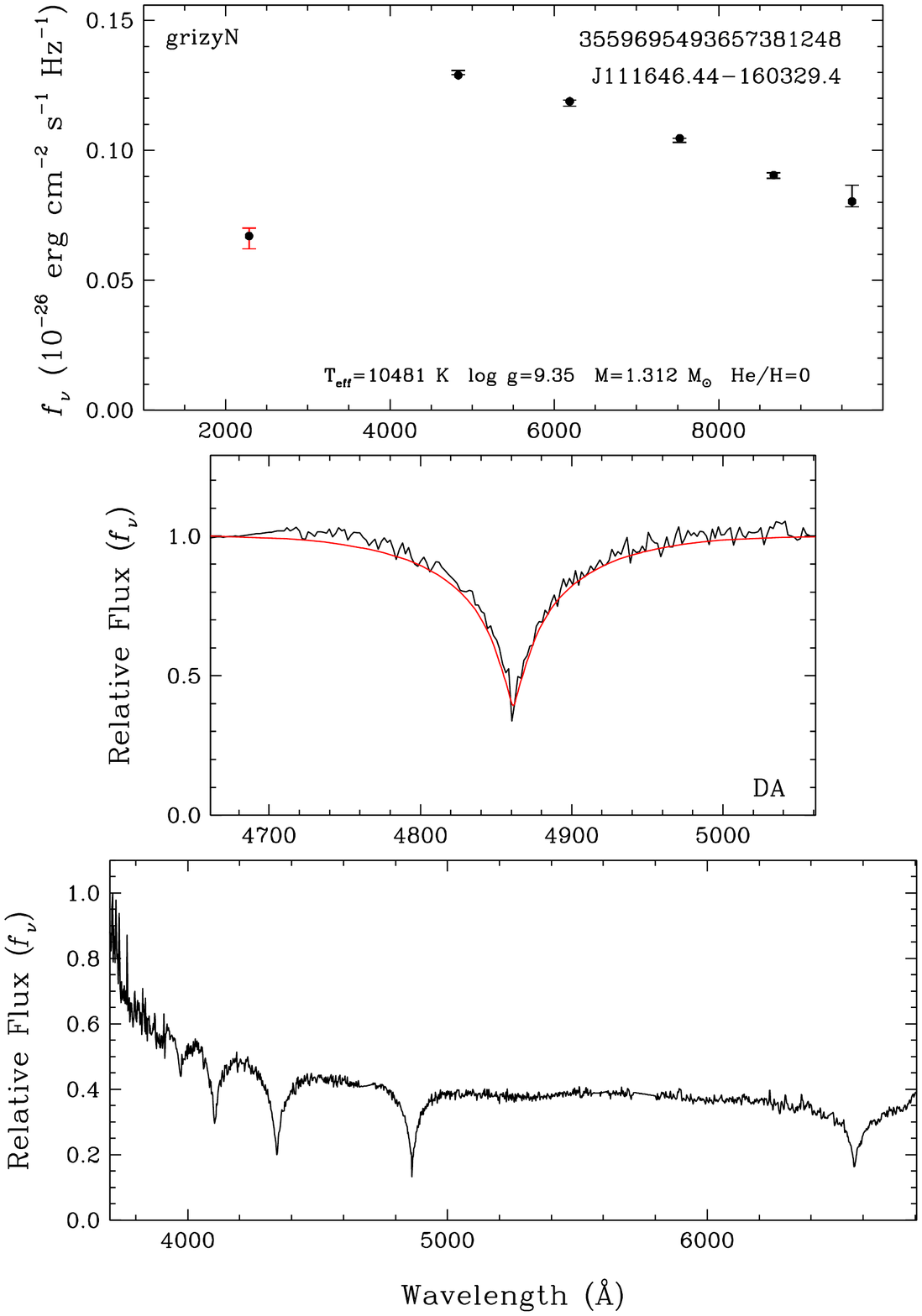}
\includegraphics[width=2.05in]{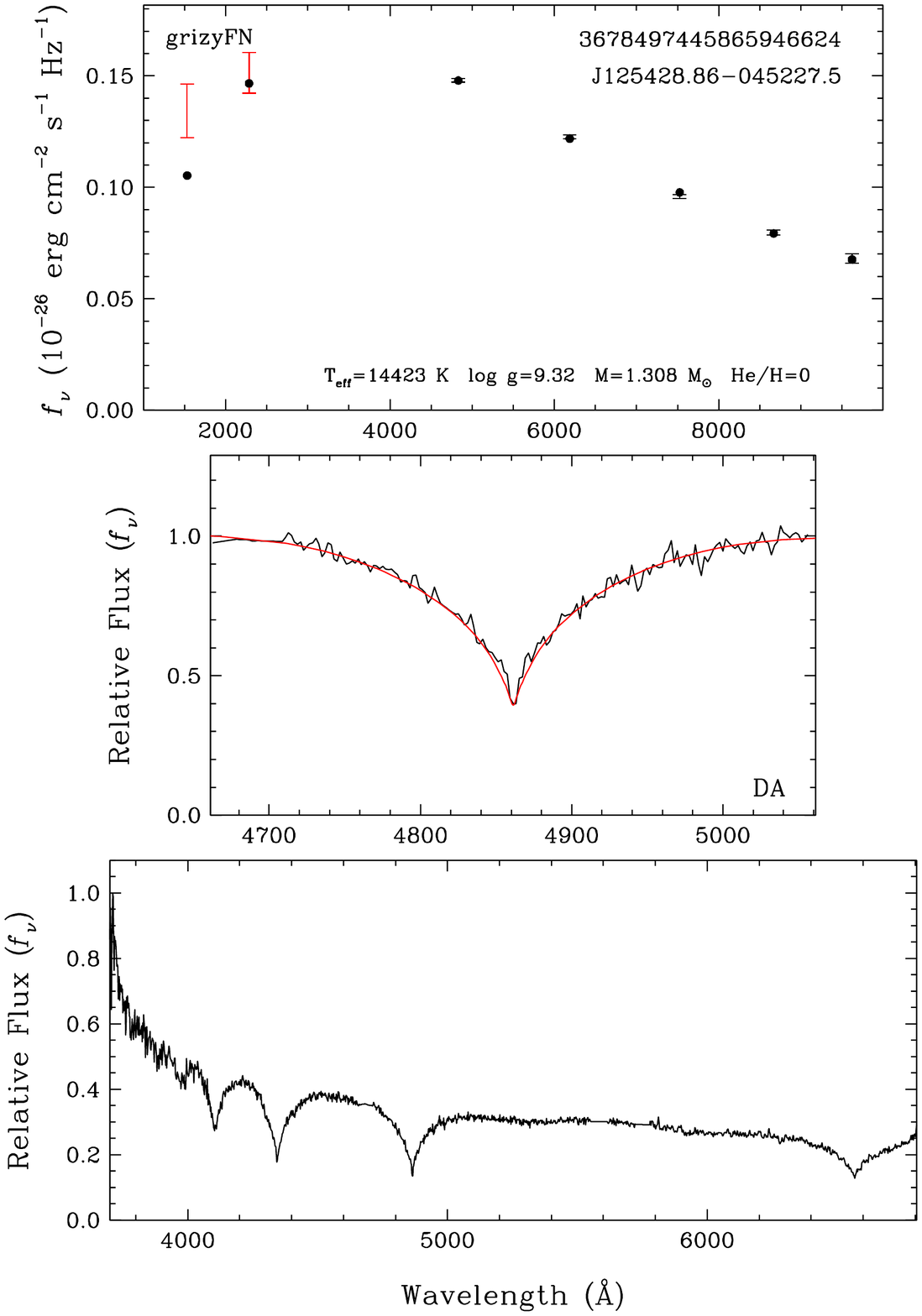}
\includegraphics[width=2.05in]{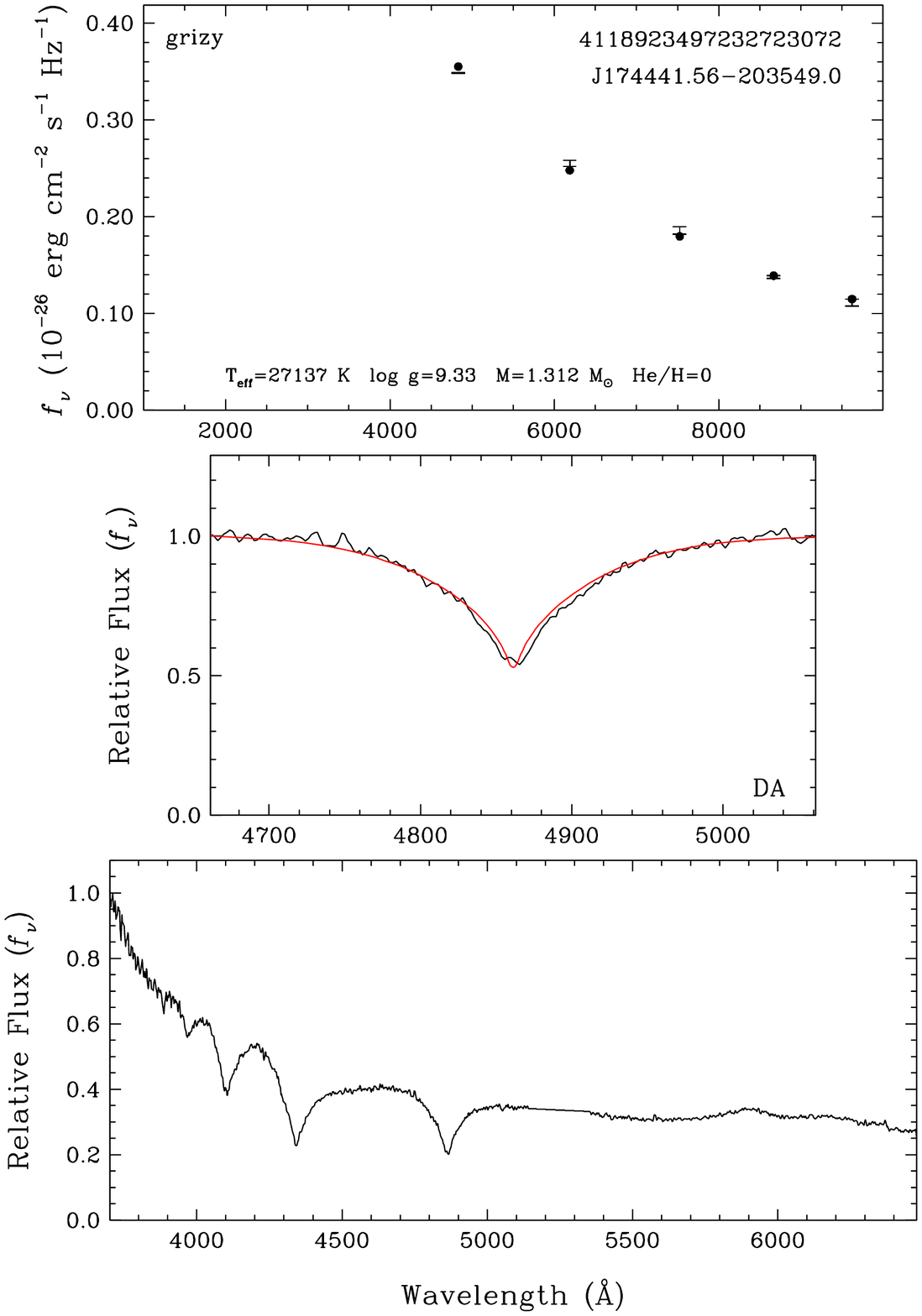}
\includegraphics[width=2.05in]{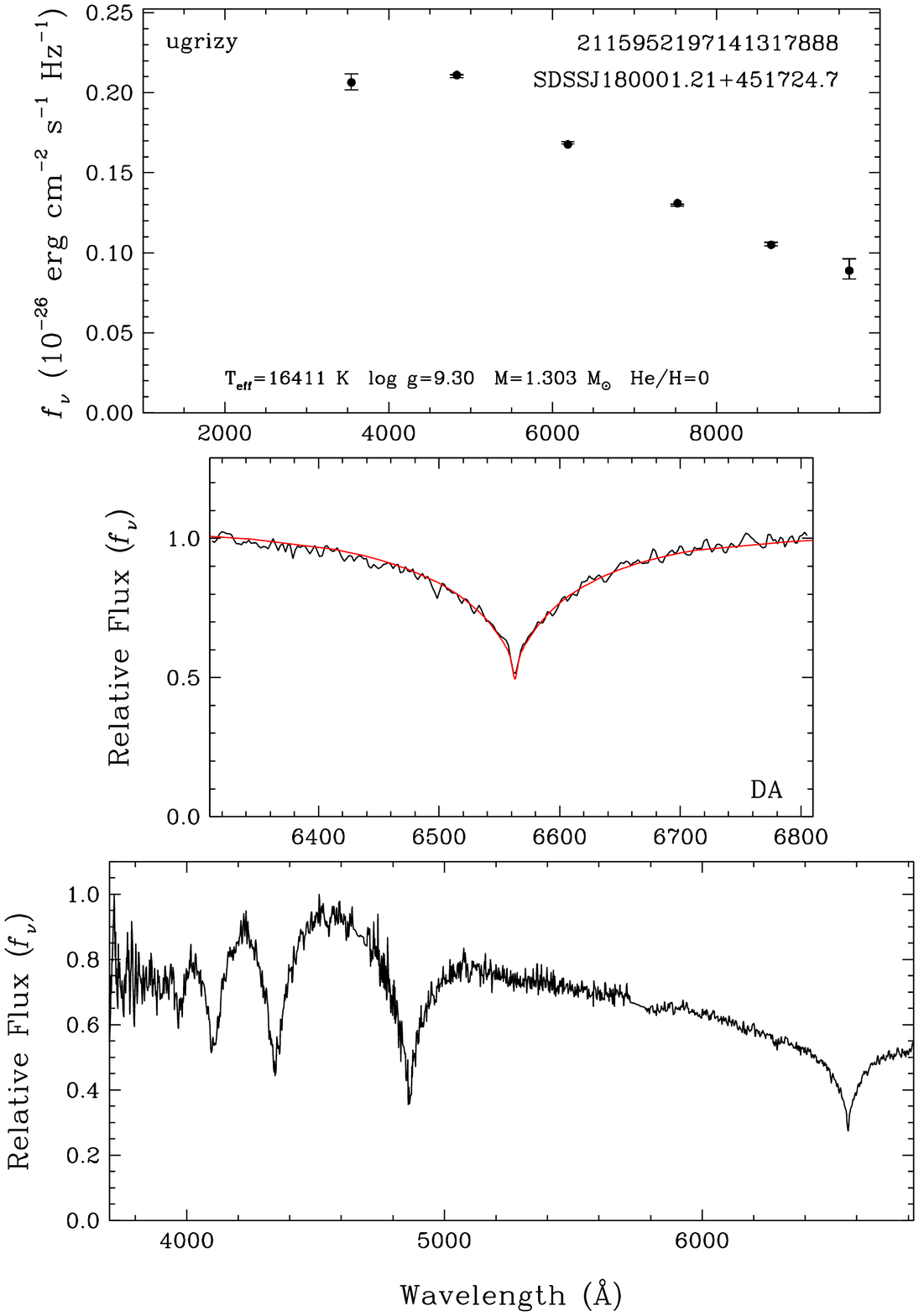}
\caption{Model atmosphere fits to eight ultramassive DA white dwarfs observed at Gemini. The top panels show
the best-fitting H (filled dots) and He (open circles) atmosphere white dwarf models to the photometry (error bars),
and includes the Gaia DR2 Source ID, object name, and the photometry used in the fitting: $ugrizyFN$ means SDSS $u$ + Pan-STARRS
$grizy$, and Galex FUV and NUV. The atmospheric parameters of the favored solution are highlighted in red. Here, and in the following
figures, we show the model parameters for CO core white dwarfs. The middle panels show the observed spectrum (black line)
along with the predicted spectrum (red line) based on the pure H atmosphere solution. The bottom panels show a broader
wavelength range for each object.}
\label{figda}
\end{figure*}

Figure \ref{figda} shows our model fits for eight ultramassive DA white dwarfs observed at Gemini. For each star, the top panel shows
the available SDSS $u$, Pan-STARRS $grizy$, and GALEX FUV and NUV photometry (error bars) along with the predicted
fluxes from the best-fitting pure hydrogen atmosphere models (filled dots). The labels in
the same panel give the Gaia DR2 Source ID, object name, and the photometry used in the fitting. The middle panel shows
the predicted spectrum based on the pure hydrogen solution, along with the observed H$\alpha$ or H$\beta$ line profiles.

We simply over-plot the predicted hydrogen line profile (red line) from the photometric fit to see if a given spectrum
is consistent with a pure hydrogen atmosphere composition, and in all cases here that is the case.
The bottom panel shows the entire Gemini GMOS spectrum of each object.
We confirm seven of these objects as ultramassive DA white dwarfs with $T_{\rm eff}>10,000$ K and $M \geq 1.3~M_{\odot}$, assuming
a CO core. The exception is J0556+1306, which is best explained by a pure H atmosphere white dwarf with $T_{\rm eff} = 8340 \pm 260$ K
and $M= 1.257 \pm 0.023~M_{\odot}$.

\begin{table*}
\centering
\caption{Physical Parameteres of our Ultramassive White Dwarf Sample assuming ONe or CO cores. All solutions
above 1.29 $M_{\odot}$ for ONe core models and above 1.334 $M_{\odot}$ for CO core models are extrapolated.
The last 10 objects had spectroscopy available in the literature prior to this work.}
\begin{tabular}{ccccccccc}
\hline
              &                    &                 &      &       ONe core & ONe  core & CO core & CO core & \\
Object  & Comp &  Spectral & $T_{\rm eff}$ &  Mass &  Cooling Age & Mass & Cooling Age & Merger\\
        &             &   Type & (K)         & ($M_{\odot}$) &   (Gyr)  & ($M_{\odot}$) & (Gyr)  & ?\\
\hline
J004917.14$-$252556.81  & H      &  DA & 13020 $\pm$ 460 & 1.263 $\pm$ 0.011 & 1.94 $\pm$ 0.08 & 1.312 $\pm$ 0.010 & 1.72 $\pm$ 0.09 & \\
J032900.79$-$212309.24  & H      &  DAH & 10330 $\pm$ 290 & 1.305 $\pm$ 0.010 & 2.32 $\pm$ 0.06 & 1.344 $\pm$ 0.008 & 1.87 $\pm$ 0.09 & \checkmark \\
J042642.02$-$502555.21  & H      &  DAH & 17900 $\pm$ 1570 & 1.264 $\pm$ 0.019 & 1.30 $\pm$ 0.16 & 1.312 $\pm$ 0.016 & 1.08 $\pm$ 0.16 & \checkmark\\
J043952.72+454302.81  & H      &  DA &19120 $\pm$ 630 & 1.258 $\pm$ 0.008 & 1.18 $\pm$ 0.06 & 1.307 $\pm$ 0.007 & 0.96 $\pm$ 0.06\\
J055631.17+130639.78  & H      &  DA & 8340 $\pm$ 260 & 1.207 $\pm$ 0.021 & 3.33 $\pm$ 0.12 & 1.257 $\pm$ 0.023 & 3.34 $\pm$ 0.18\\
J060853.60$-$451533.03  & H      &  DC/DAH & 19580 $\pm$ 1910 & 1.258 $\pm$ 0.021 & 1.13 $\pm$ 0.16 & 1.307 $\pm$ 0.019 & 0.92 $\pm$ 0.17 & \checkmark\\
J070753.00+561200.25  & H      &  DC/DAH & 18100 $\pm$ 350 & 1.240 $\pm$ 0.005 & 1.23 $\pm$ 0.04 & 1.291 $\pm$ 0.005 & 1.06 $\pm$ 0.04 & \checkmark\\
J080502.93$-$170216.57  & H      &  DAH & 10830 $\pm$ 110 & 1.254 $\pm$ 0.004 & 2.40 $\pm$ 0.03 & 1.304 $\pm$ 0.003 & 2.20 $\pm$ 0.03 & \checkmark\\
\dots                        & [H/He]=$-5$ & \dots & 10010 $\pm$ 120 & 1.213 $\pm$ 0.004 & 2.70 $\pm$ 0.04 & 1.249 $\pm$ 0.006 & 2.23 $\pm$ 0.04\\
J093430.71$-$762614.48  & H      &  DAH & 10050 $\pm$ 1350 & 1.284 $\pm$ 0.055 & 2.47 $\pm$ 0.35 & 1.328 $\pm$ 0.047 & 2.11 $\pm$ 0.50 & \checkmark\\
\dots                        & [H/He]=$-5$ & \dots &   9180 $\pm$ 1050 & 1.238 $\pm$ 0.052 & 2.86 $\pm$ 0.33 & 1.279 $\pm$ 0.051 & 2.32 $\pm$ 0.46\\
J095933.33$-$182824.16  & H      &  DA & 12000 $\pm$ 180 & 1.273 $\pm$ 0.005 & 2.12 $\pm$ 0.03 & 1.320 $\pm$ 0.004 & 1.83 $\pm$ 0.04\\
J111646.44$-$160329.42  & H      &  DA & 10480 $\pm$ 170 & 1.264 $\pm$ 0.007 & 2.45 $\pm$ 0.05 & 1.312 $\pm$ 0.006 & 2.21 $\pm$ 0.07 & \checkmark\\
J125428.86$-$045227.48  & H      &  DA & 14420 $\pm$ 390 & 1.258 $\pm$ 0.008 & 1.71 $\pm$ 0.06 & 1.308 $\pm$ 0.007 & 1.52 $\pm$ 0.06\\
J174441.56$-$203549.05  & H      &  DA & 27140 $\pm$ 890 & 1.271 $\pm$ 0.008 & 0.65 $\pm$ 0.06 & 1.312 $\pm$ 0.008 & 0.43 $\pm$ 0.04\\
J180001.21+451724.7   & H      &  DA & 16410 $\pm$ 290 & 1.253 $\pm$ 0.003 & 1.44 $\pm$ 0.03 & 1.303 $\pm$ 0.004 & 1.26 $\pm$ 0.04\\
J181913.36$-$120856.44  & C/O    &  hotDQ & 23800  & \dots & \dots & 1.243 & 0.42 & \checkmark\\
\hline
J010338.56$-$052251.96 & H & DAH: & 9040 $\pm$ 70 & 1.262 $\pm$ 0.003 & 2.84 $\pm$ 0.03 & 1.310 $\pm$ 0.003 & 2.60 $\pm$ 0.04 & \checkmark\\
J025431.45+301935.38 & [H/He]=$-5$ & DC & 11060 $\pm$ 560 & 1.302 $\pm$ 0.024 & 2.25 $\pm$ 0.10 & 1.330 $\pm$ 0.016 & 1.49 $\pm$ 0.17\\
\dots                               & [C/He]=$-4$ & \dots & 10190 $\pm$ 290 & 1.261 $\pm$ 0.016 & 2.53 $\pm$ 0.08  & 1.301 $\pm$ 0.014 & 1.93 $\pm$ 0.12\\
J114012.81+232204.7 & H & DA & 11860 $\pm$ 220 & 1.294 $\pm$ 0.008 & 2.10 $\pm$ 0.04 & 1.336 $\pm$ 0.006 & 1.71 $\pm$ 0.06 \\
J132926.04+254936.4 & H & DA & 29010 $\pm$ 750 & 1.314 $\pm$ 0.006 & 0.81 $\pm$ 0.05 & 1.351 $\pm$ 0.006 & 0.37 $\pm$ 0.03 \\
J172736.28+383116.9 & H & DA & 9420 $\pm$ 200 & 1.252 $\pm$ 0.012 & 2.78 $\pm$ 0.08  & 1.302 $\pm$ 0.011 & 2.59 $\pm$ 0.12 \\
J183202.83+085636.24 & He & DBA & 34210 $\pm$ 1020 & 1.301 $\pm$ 0.006 & 0.45 $\pm$ 0.03 & 1.319 $\pm$ 0.004 & 0.20 $\pm$ 0.02 & \checkmark\\
J190132.74+145807.18 & H & DC/DAH & 29100 $\pm$ 480 & 1.279 $\pm$ 0.003 & 0.61 $\pm$ 0.02 & 1.319 $\pm$ 0.004 & 0.35 $\pm$ 0.02 & \checkmark\\
J221141.80+113604.5 & [H/He]=$-1.5$ & DAH & 7500 - 8390 & 1.231 $\pm$ 0.010 &  3.1 - 3.2  & 1.268 $\pm$ 0.010 &  2.6 - 2.9  & \checkmark\\
J225513.48+071000.9 & H & DC/DAH & 10990 $\pm$ 210 & 1.252 $\pm$ 0.012 & 2.36 $\pm$ 0.05 & 1.302 $\pm$ 0.011 & 2.18 $\pm$ 0.09 & \checkmark\\
J235232.30$-$025309.2 & H & DA & 10680 $\pm$ 100 & 1.272 $\pm$ 0.003 & 2.38 $\pm$ 0.02 & 1.319 $\pm$ 0.003 & 2.10 $\pm$ 0.03 & \checkmark \\
\hline
\end{tabular}
\end{table*}

\begin{figure*}
\centering
\includegraphics[width=3.2in, angle=-90, clip=true, trim=3.2in 0.8in 0.8in 0.8in]{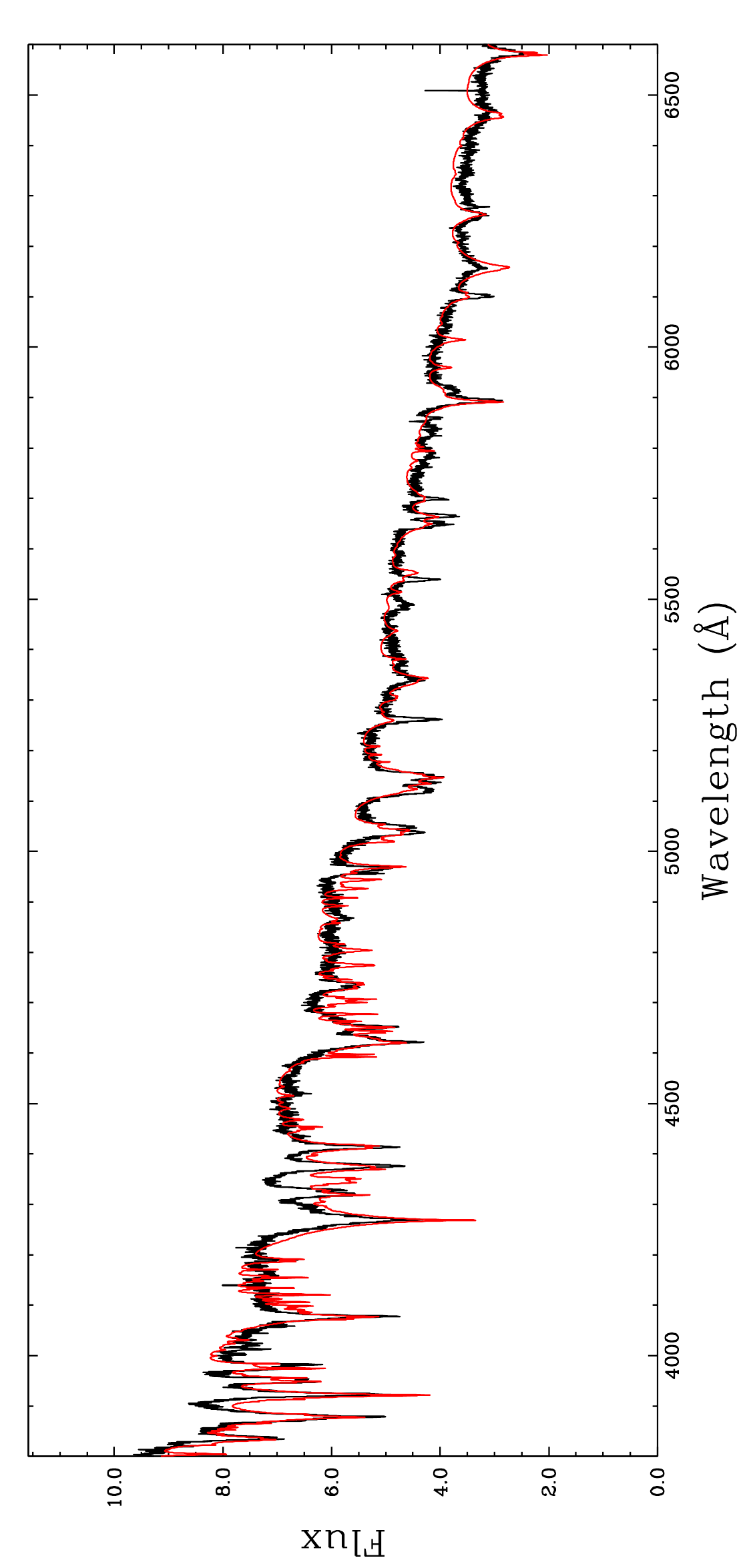}
\caption{Model atmosphere fit to the hot DQ white dwarf J1819$-$1208 assuming equal amounts of carbon and oxygen in the
atmosphere. The best-fitting effective temperature is 23800 K.}
\label{figdq}
\end{figure*}

\subsection{A Hot DQ White Dwarf}

Figure \ref{figdq} shows our model fits to J1819$-$1208, a unique object in our sample with no traces of hydrogen or helium lines in its spectrum.
In fact, the optical spectrum of J1819$-$1208 is dominated by carbon and oxygen lines, making it a member of the hot DQ spectral type.
Hot DQ white dwarfs with temperatures between about 18,000 and 24,000 K are unique in having atmospheres dominated by carbon and oxygen
\citep{dufour07,dufour08}. 

We rely on a new model atmosphere grid for hot DQ stars \citep[see][]{dufour11} with effective temperatures between 16,000 and 25,000 K
for a fixed value of C/O = 1.0 for this exploratory study. The details of these models will be presented in a future publication. We fixed the
surface gravity to $\log{g}=9.0$ and fitted for the effective temperature. The best-fitting model has $T_{\rm eff} = 23,800$ K and is
shown in red in Figure \ref{figdq}. This model does a decent job of matching the spectral features in the spectrum of J1819$-$1208, though
some of the oxygen lines are weaker than predicted by this model, indicating that the oxygen to carbon ratio is likely smaller than one.  
We defer a detailed model atmosphere analysis of this object to a future publication, but confirm that J1819$-$1208 is a relatively hot DQ white 
dwarf with a carbon and oxygen atmosphere. Fixing the effective temperature at 23,800 K, the spectral energy distribution
of J1819$-$1208 based on Pan-STARRS photometry and Gaia parallax indicates a mass of $1.24~M_{\odot}$ and a cooling age of 420 Myr.

\subsection{Magnetic White Dwarfs}

\begin{figure}
\centering
\includegraphics[width=3.4in, clip=true, trim=0.2in 0.2in 0.7in 0.6in]{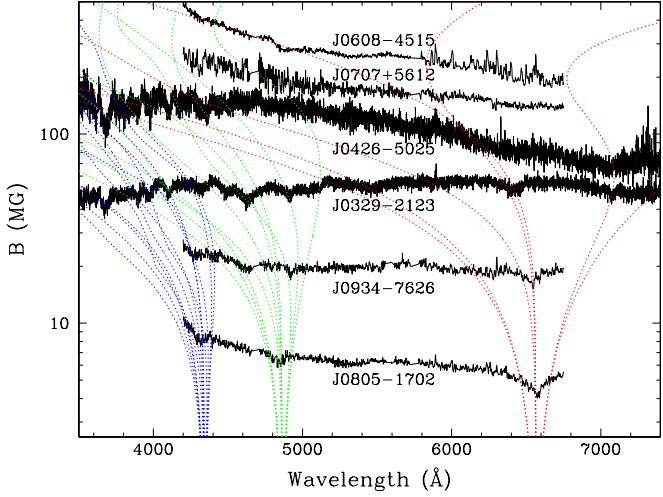}
\caption{Spectra of six newly identified magnetic white dwarfs compared to predicted line positions
of H$\alpha$, H$\beta$, and H$\gamma$ as a function of the magnetic field strength \citep{schimeczek14}.
The bottom three objects, J0805$-$1702, J0934$-$7626, and J0329$-$2123 have $B\sim$5, 25, and 50 MG,
respectively. The remaining three objects are strongly magnetic, but it is difficult to
constrain their field strength based on the available data.}
\label{figmag}
\end{figure}

Six of the newly observed ultramassive white dwarfs are magnetic. Figure \ref{figmag} shows the 
Gemini and Magellan spectra of these six targets along with the predicted Zeeman components of H$\alpha$, H$\beta$,
and H$\gamma$ as a function of the magnetic field strength from \citet{schimeczek14}.
Two of these targets, J0805$-$1702 and J0934$-$7626 show H$\alpha$ near its rest wavelength, and are compatible
with $B\sim5-25$ MG fields (see below). Two other targets, J0329$-$2123 and J0426$-$5025, show several absorption features
in the blue that likely require $B\sim50-100$ MG fields. Yet two other targets, J0608$-$4515 and
J0707+5612, show essentially featureless spectra, but have effective temperatures near
18,000 - 20,000 K based on their overall spectral energy distributions. Hence, the only way for them to have featureless
spectra is if they are strongly magnetic. Fitting the spectra of these strongly magnetic white dwarfs is beyond the scope
of this paper, and their location in Figure \ref{figmag} is arbitrary and should not be taken as an indication of their actual
magnetic field strengths.

For J0805$-$1702 and J0934$-$7626, the two magnetic white dwarfs where H$\alpha$ and H$\beta$ are clearly visible,
we computed magnetic model spectra using an approach similar to that described in \citet{bergeron92} and \citet{kilic21b}.
We use offset dipole models, where the independent parameters are the dipole field strength $B_d$, the dipole offset $a_z$
measured in units of stellar radius from the center of the star, and the viewing angle $i$ between the dipole axis and the line
of sight ($i=0^{\circ}$ for a pole-on view).

\begin{figure}
\centering
\includegraphics[width=3.4in, clip=true, trim=0.3in 3.6in 0.6in 3.3in]{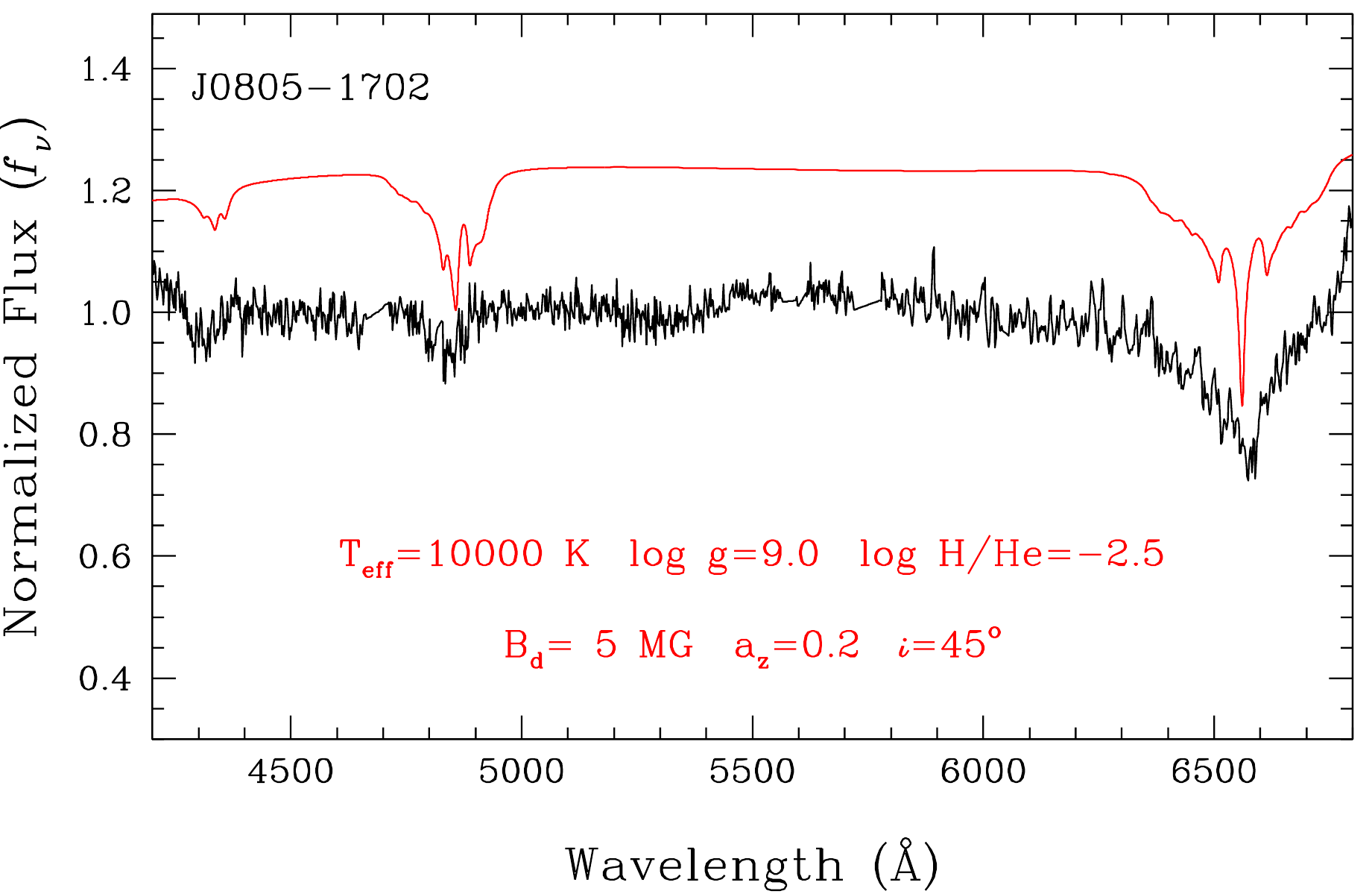}
\includegraphics[width=3.4in, clip=true, trim=0.3in 3.6in 0.6in 3.3in]{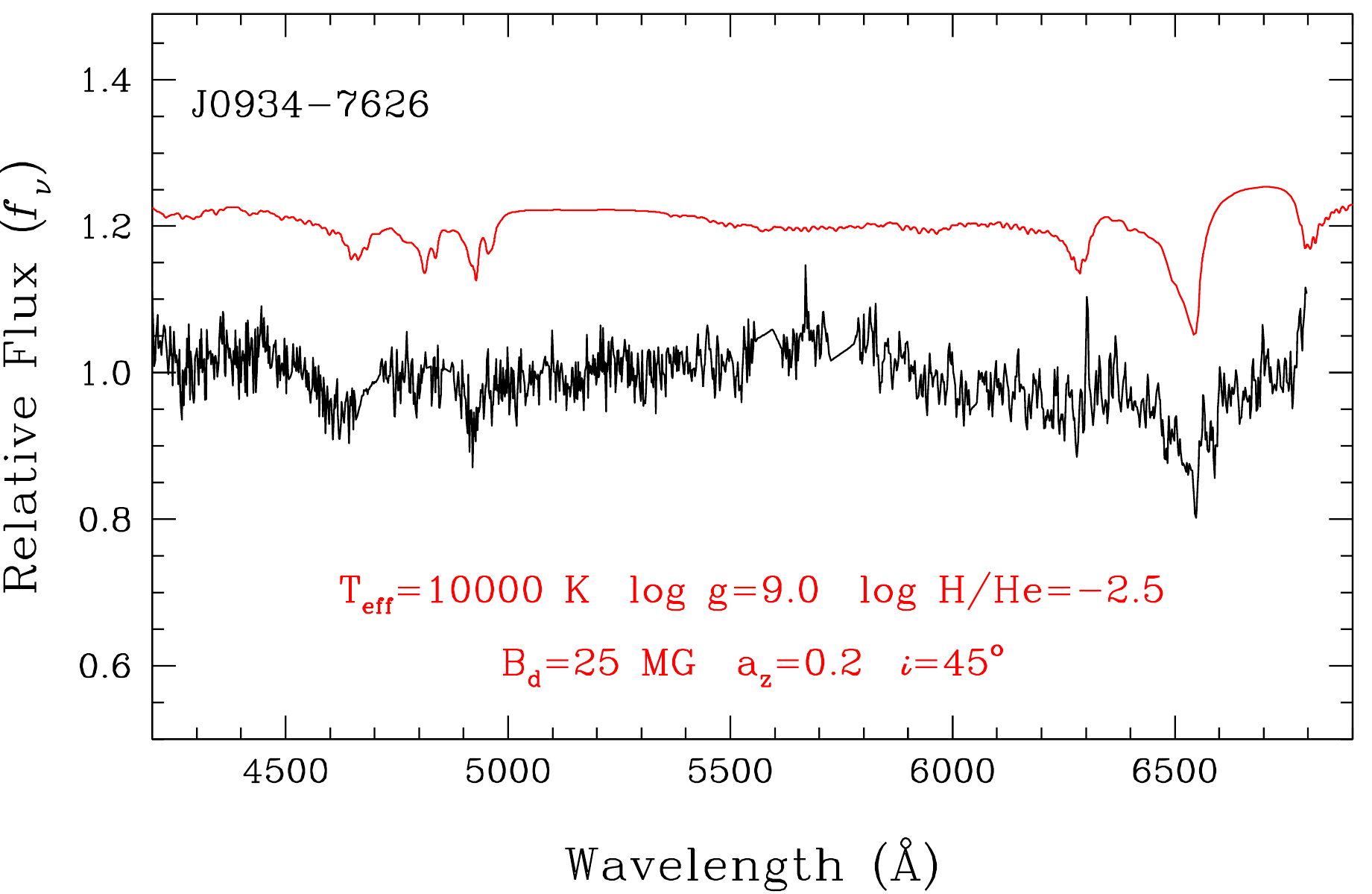}
\caption{A comparison between the observed Gemini spectra of the magnetic DA white dwarfs J0805$-$1702
and J0934$-$7626 and a mixed H/He atmosphere white dwarf model with
$T_{\rm eff}=10,000$ K, $\log{g}=9$, $\log {\rm H/He}=-2.5$, a dipole field strength of $B_d=5$ MG
(for J0805$-$1702) and 25 MG (for J0934$-$7626), the dipole offset $a_z=0.2$, and a viewing angle of $i=45^{\circ}$.}
\label{fitmag}
\end{figure}

The best-fitting models under the assumption of a pure hydrogen composition for these two stars require
the Balmer lines to be stronger than observed, regardless of the field strength and geometry. One way to reduce
the strength of the Balmer lines is if these stars have mixed hydrogen and helium atmospheres, similar to the magnetic
white dwarf J2211+1136 \citep{kilic21b}. Figure \ref{fitmag} shows a comparison of the observed spectra of 
J0805$-$1702 and J0934$-$7626 with a model where $T_{\rm eff}=10,000$ K, $\log{g}=9$, $\log {\rm H/He}=-2.5$,
$B_d=5$ MG (for J0805$-$1702) and 25 MG (for J0934$-$7626), $a_z=0.2$, and $i=45^{\circ}$. Note that
the model shown in each panel is not a fit. We simply overplot these models to demonstrate that 1) both
J0805$-$1702 and J0934$-$7626 have mixed hydrogen and helium atmospheres, and 2) they have field strengths
of $\sim$5 and 25 MG, respectively.

Table 1 presents the physical parameters of all 25 ultramassive white dwarfs in our sample.
For completeness, we provide the masses and the cooling ages for both ONe and CO core compositions.
Table 1 also includes the 10 objects with spectroscopy analyzed in \citet{kilic21,kilic21b} at the bottom, which includes 3 objects
classified as DC. However, only one of these objects, J0254$+$3019, is a genuine DC white dwarf with no clear evidence
of magnetism. The other two, J1901+1458 and J2255+0710, show broad shifted Zeeman absorption features that indicate
a strong magnetic field \citep{caiazzo21,kilic21b}. Interestingly, 10 out of the 25 white dwarfs in our sample show
evidence of magnetism in their optical spectra. This is about a factor of four higher than the fraction of magnetic white dwarfs within
the local 20 pc sample \citep{holberg16}. Note that our low-resolution spectroscopy is not sensitive to fields strengths below about
100 kG. Hence, there may be other magnetic white dwarfs with weaker fields hiding in the sample. 

\section{Variability}

\begin{figure*}
\centering
\includegraphics[width=2.3in, clip=true, trim=0.3in 2in 0.4in 1.3in]{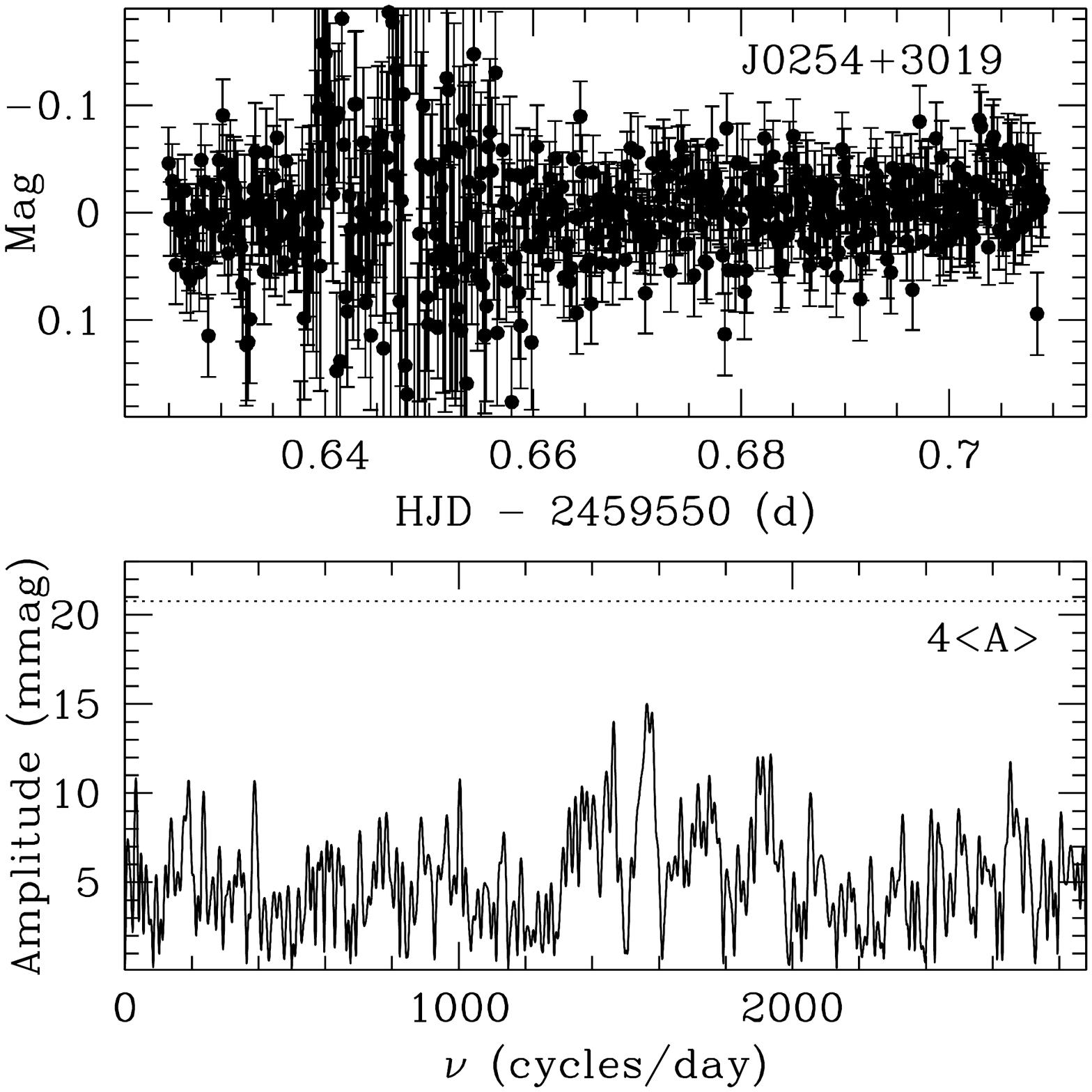}
\includegraphics[width=2.3in, clip=true, trim=0.3in 2in 0.4in 1.3in]{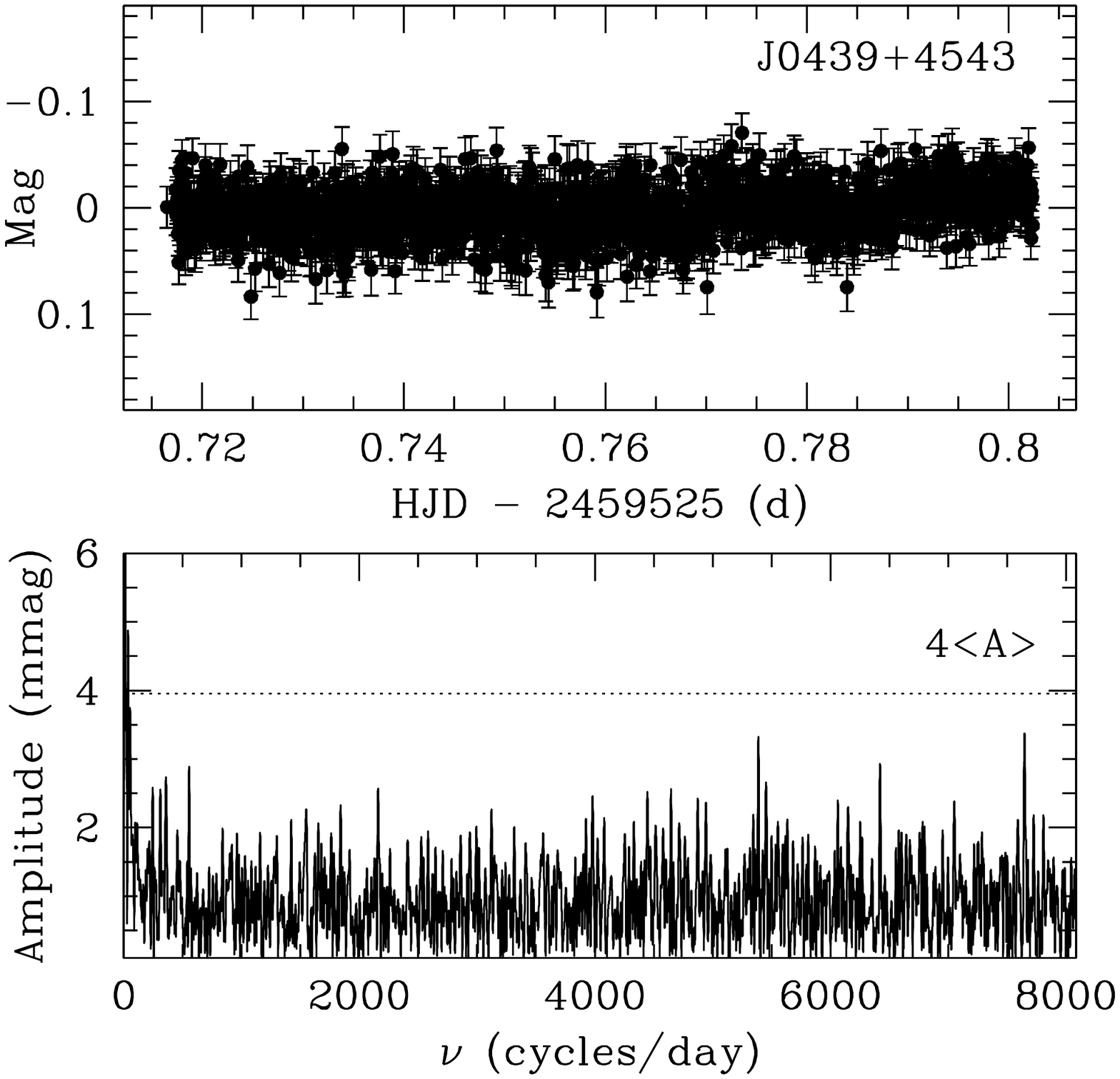}
\includegraphics[width=2.3in, clip=true, trim=0.3in 2in 0.4in 1.3in]{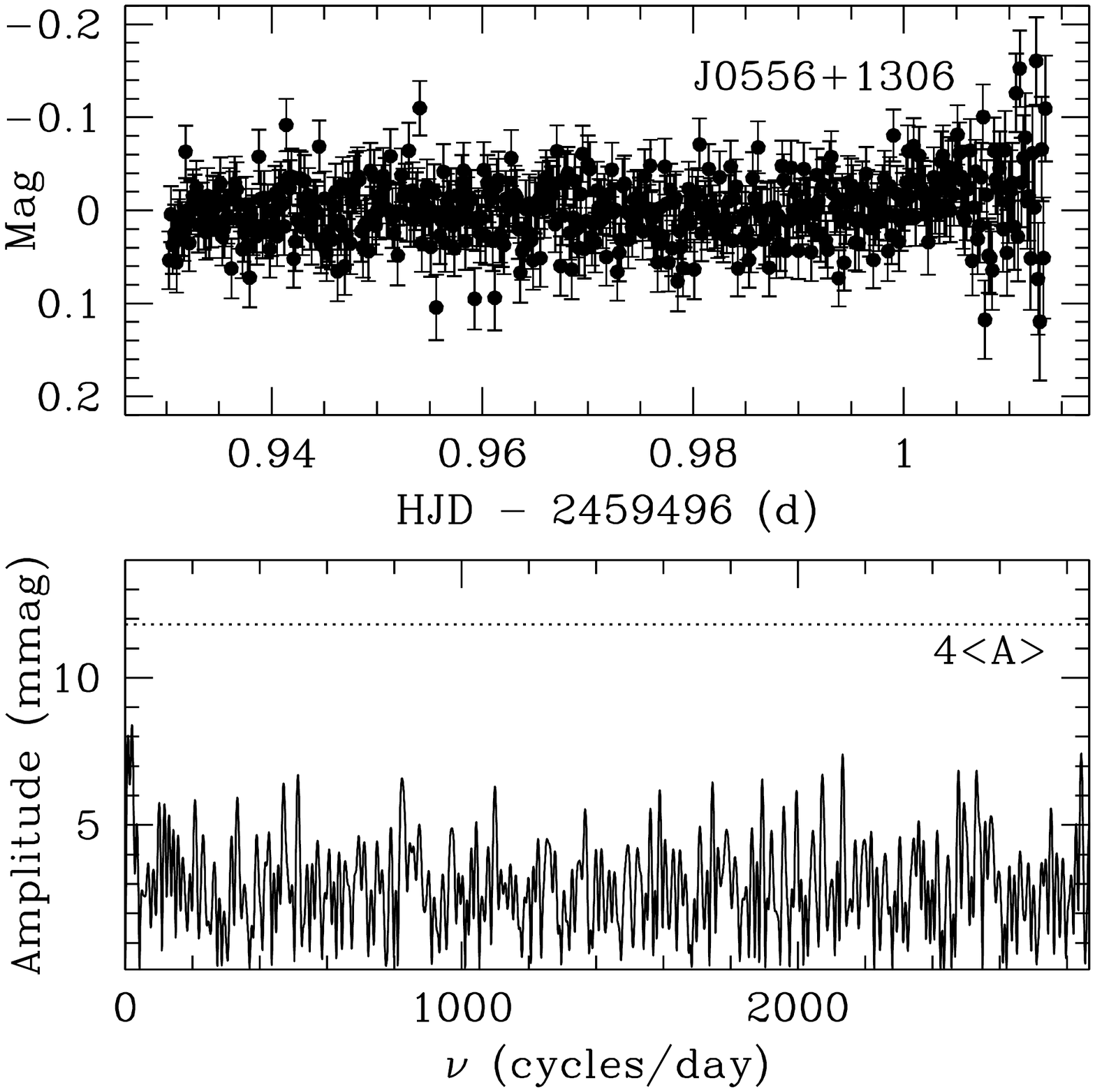}
\includegraphics[width=2.3in, clip=true, trim=0.3in 2in 0.4in 1.3in]{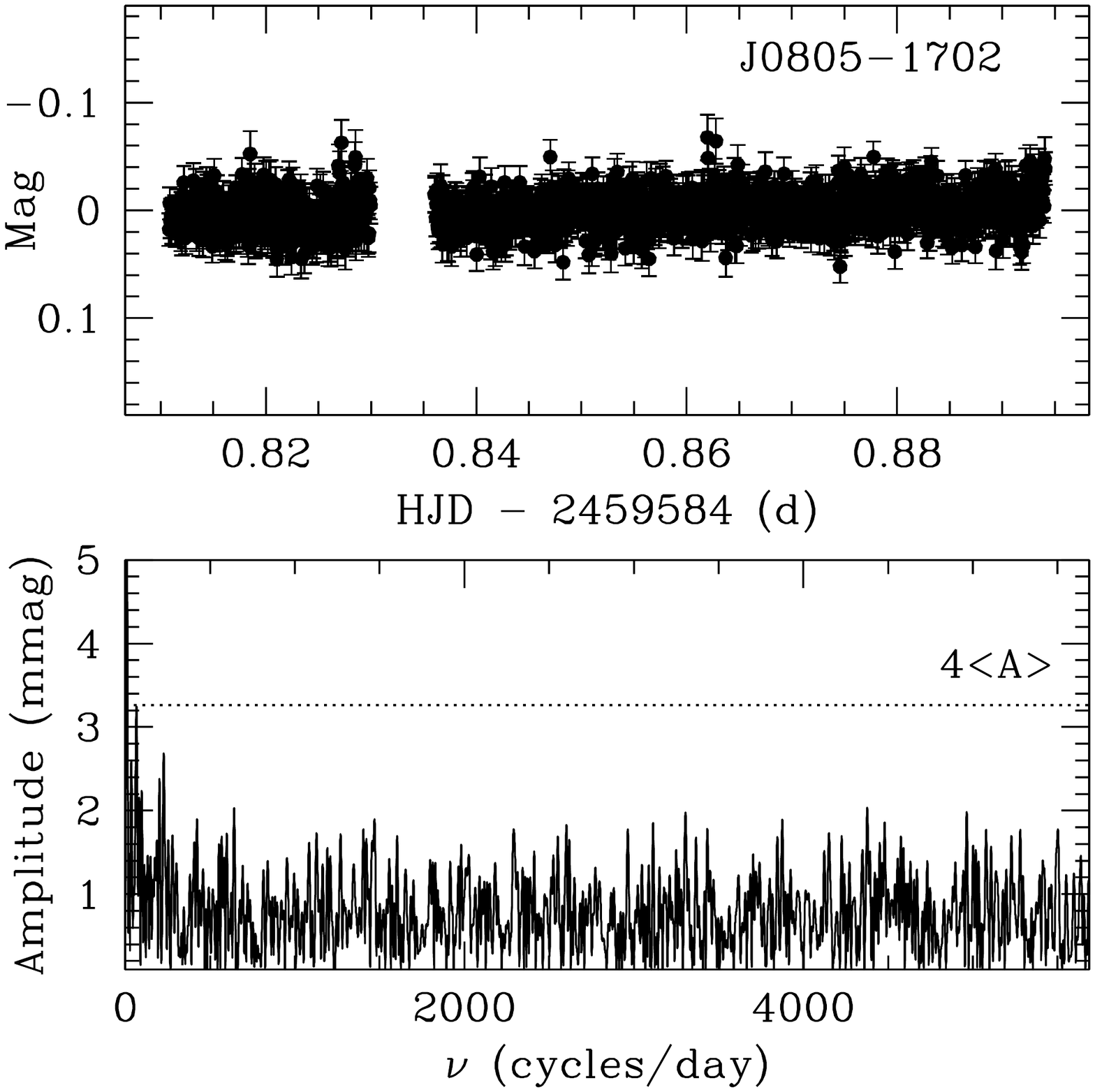}
\includegraphics[width=2.3in, clip=true, trim=0.3in 2in 0.4in 1.3in]{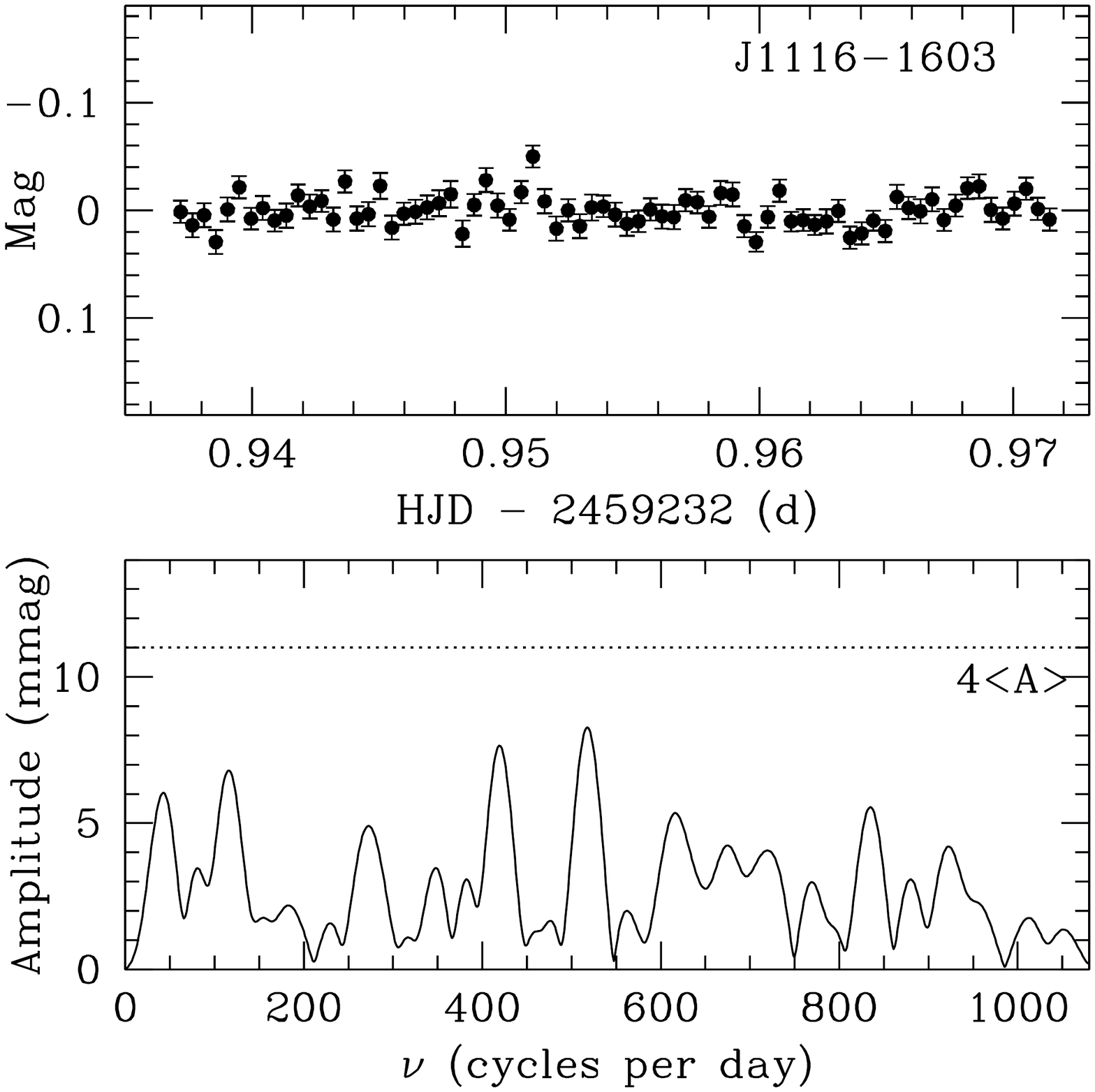}
\includegraphics[width=2.3in, clip=true, trim=0.3in 2in 0.4in 1.3in]{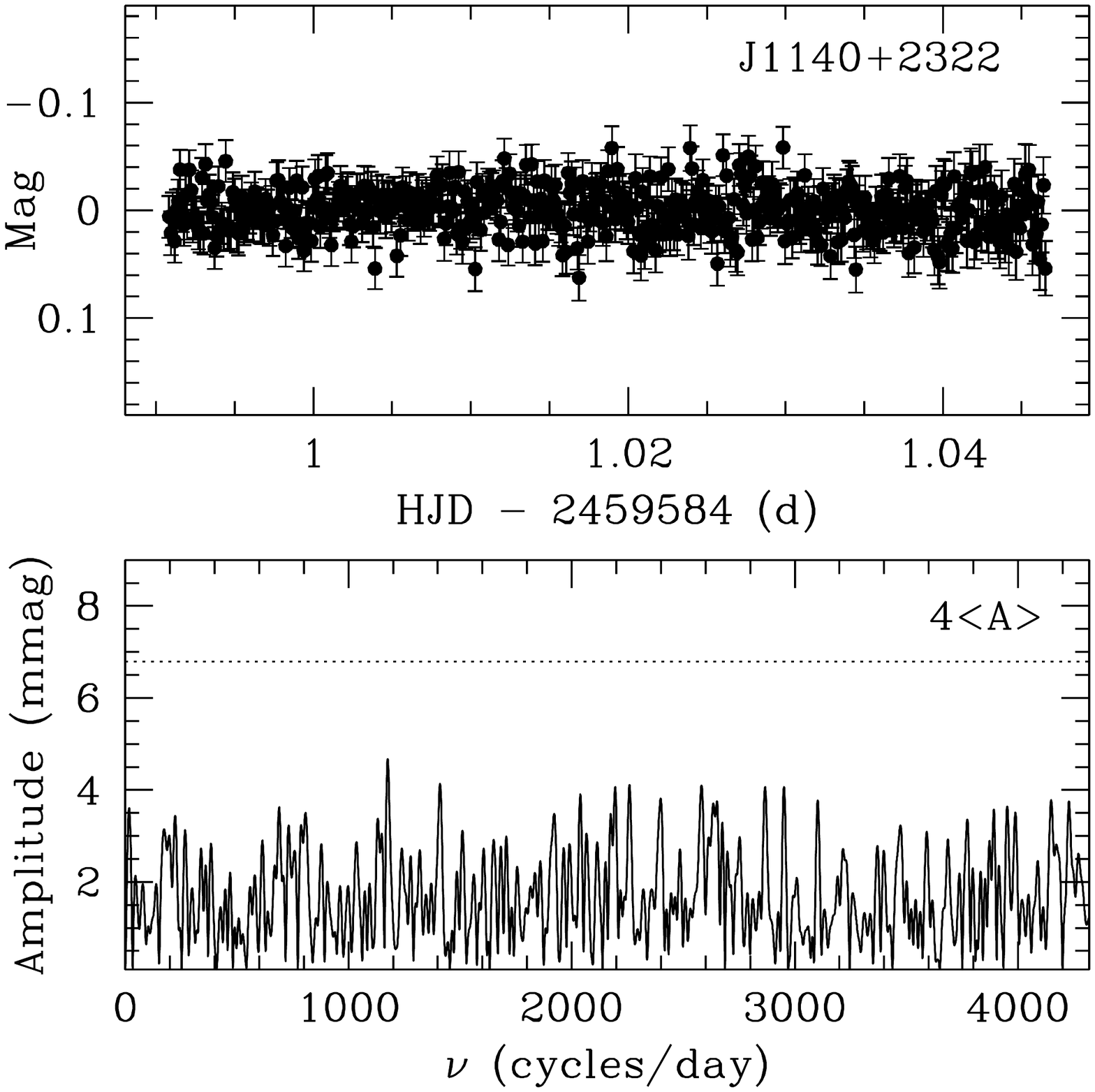}
\includegraphics[width=2.3in, clip=true, trim=0.3in 2in 0.4in 1.3in]{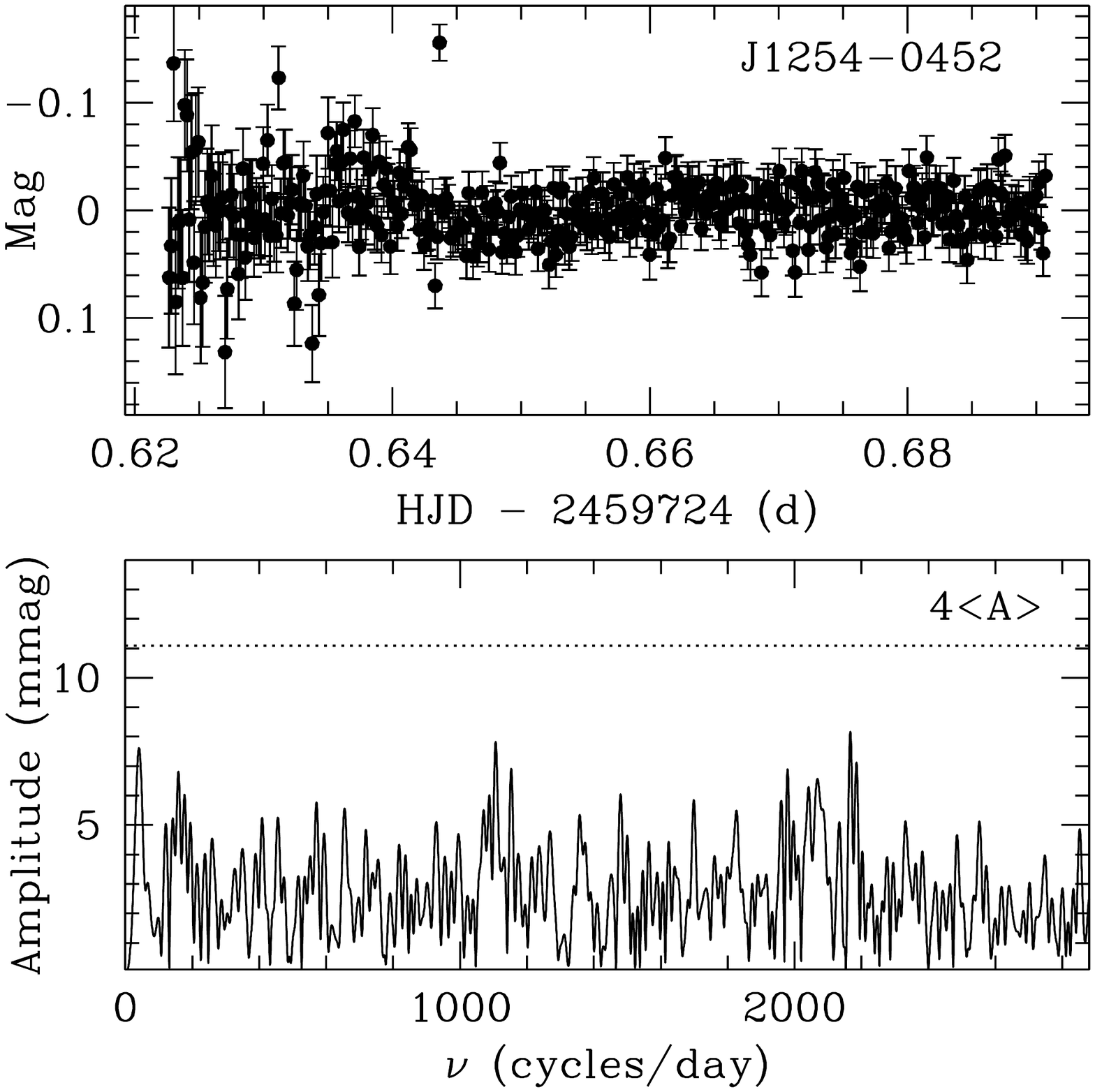}
\includegraphics[width=2.3in, clip=true, trim=0.3in 2in 0.4in 1.3in]{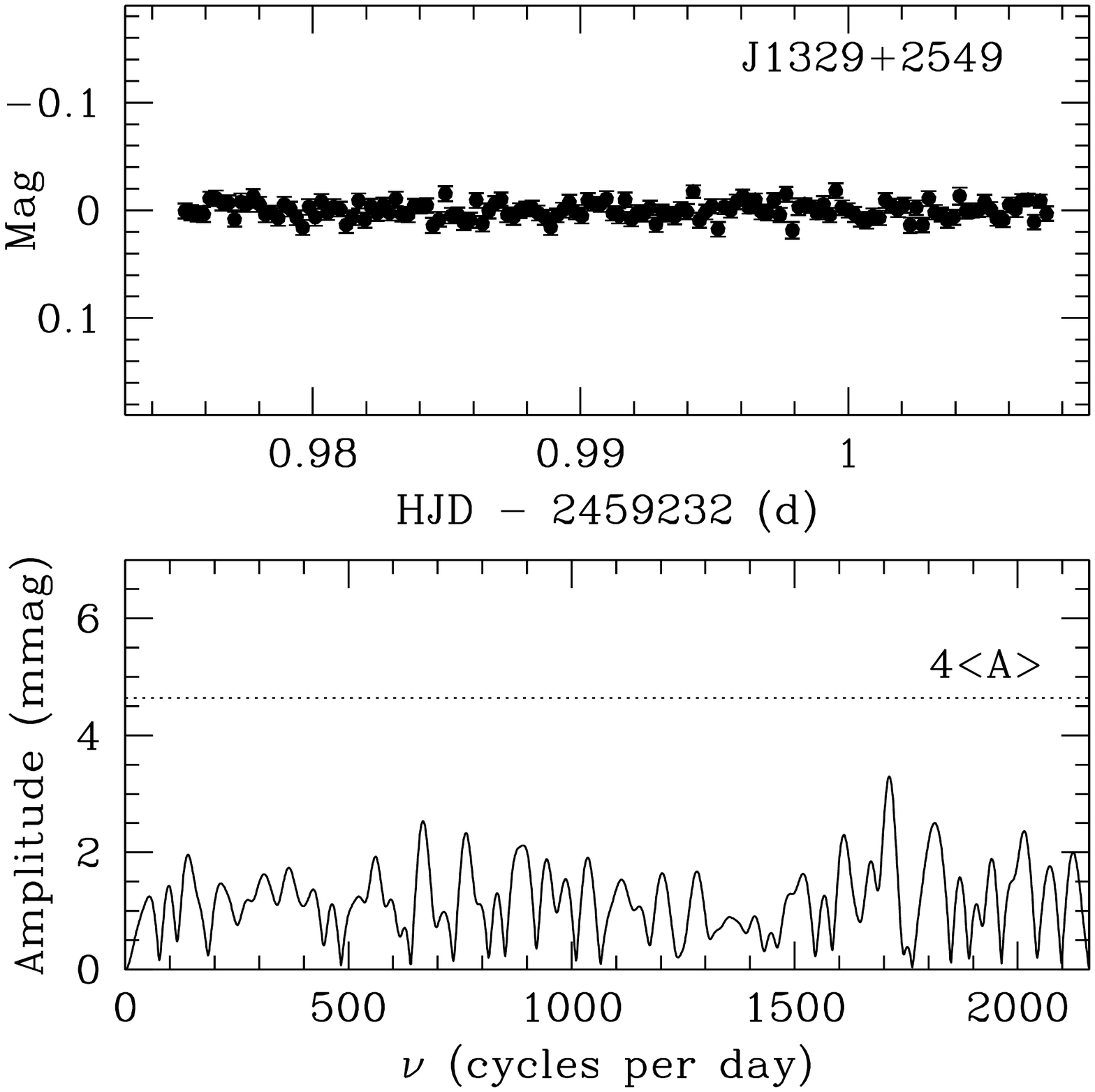}
\includegraphics[width=2.3in, clip=true, trim=0.3in 2in 0.4in 1.3in]{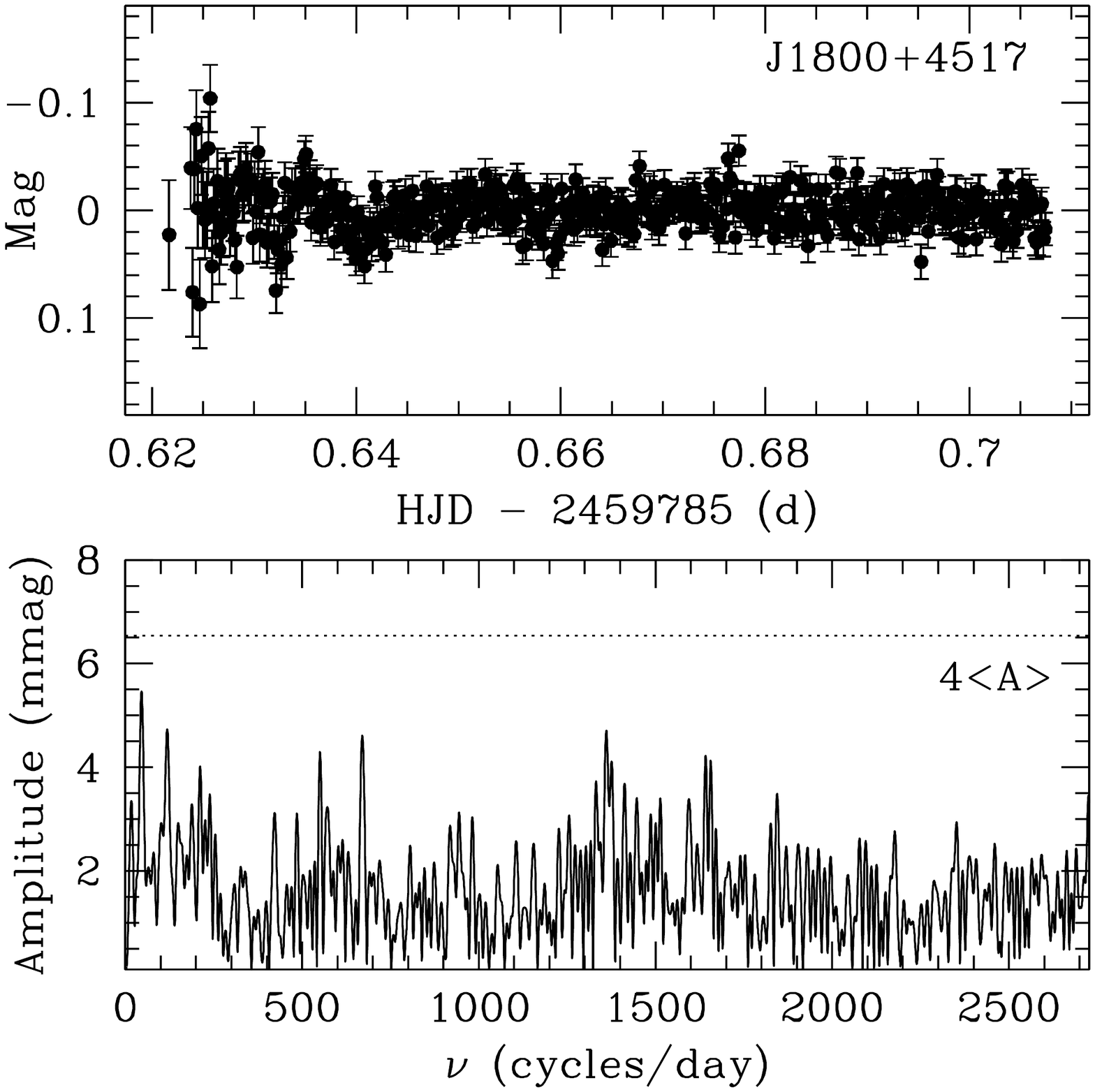}
\includegraphics[width=2.3in, clip=true, trim=0.3in 2in 0.4in 1.3in]{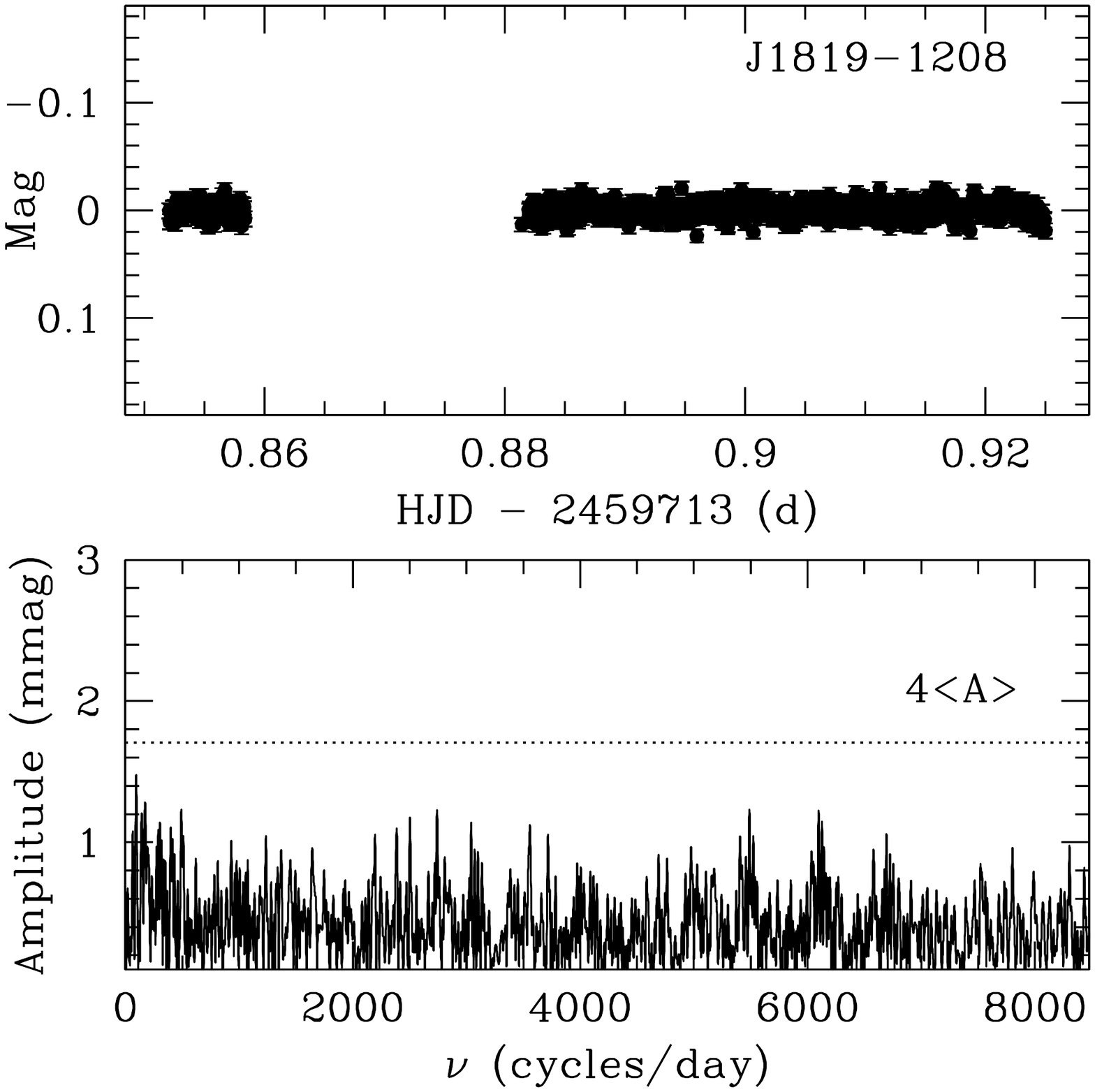}
\includegraphics[width=2.3in, clip=true, trim=0.3in 2in 0.4in 1.3in]{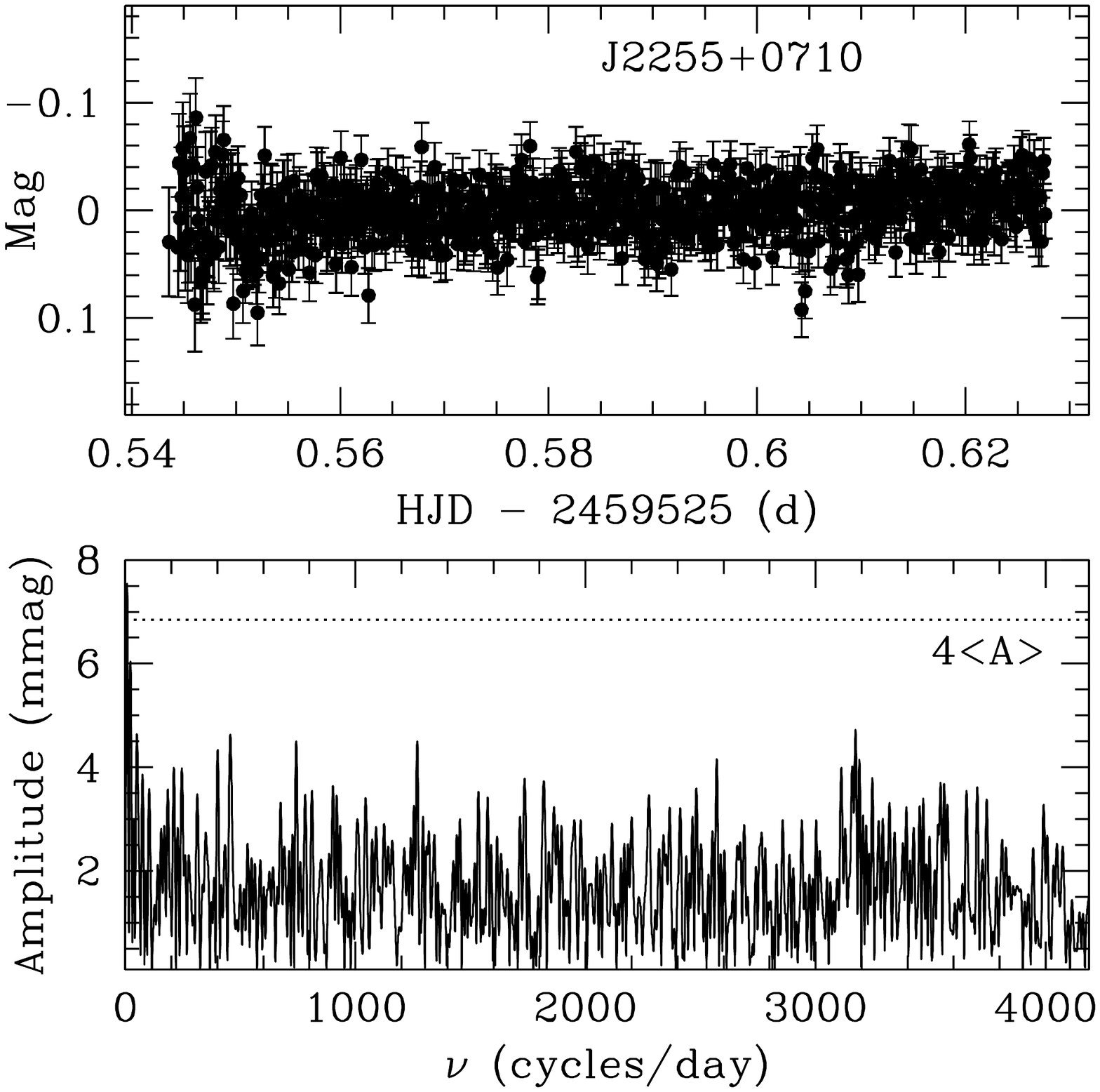}
\includegraphics[width=2.3in, clip=true, trim=0.3in 2in 0.4in 1.3in]{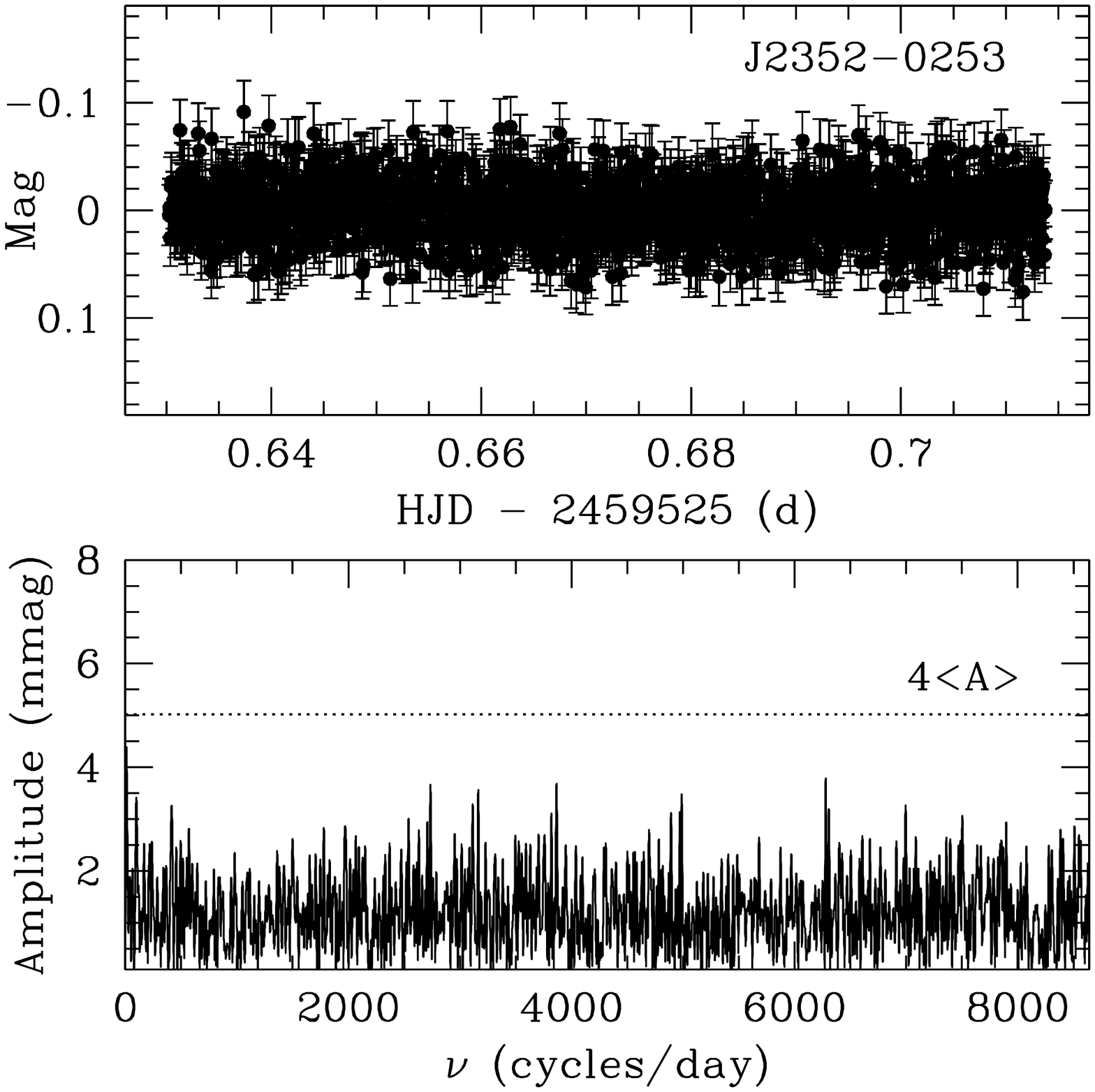}
\caption{Light curves (top) and their Fourier transforms (bottom) of 12 non-variable white dwarfs observed
at the APO 3.5m telescope. The dotted lines mark the 4$\langle {\rm A}\rangle$ level, where $\langle {\rm A}\rangle$
is the average amplitude in the Fourier transform.}
\label{figapo}
\end{figure*}

\begin{figure*}
\centering
\includegraphics[width=2.3in, clip=true, trim=0.3in 2in 0.4in 1.3in]{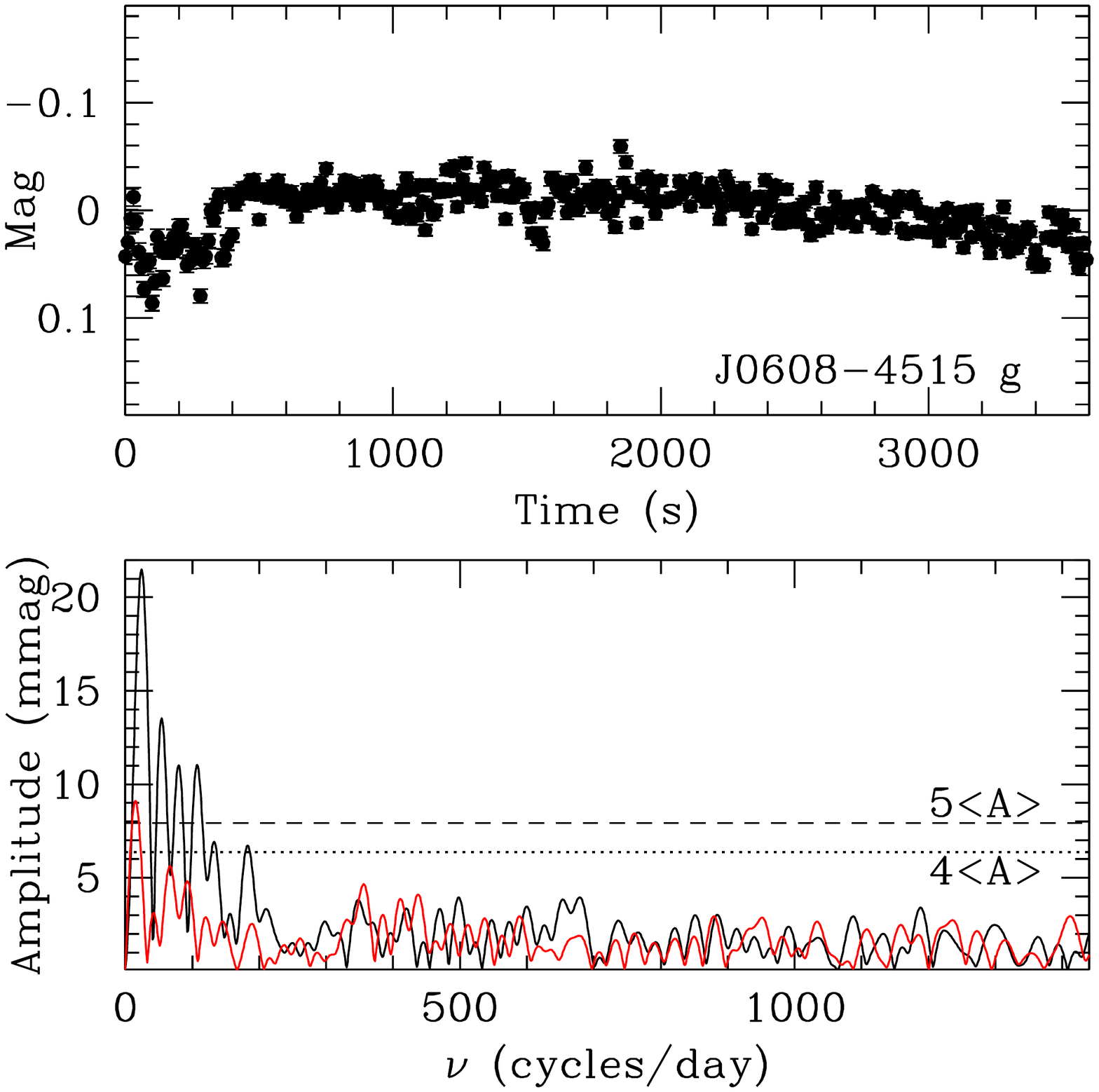}
\includegraphics[width=2.3in, clip=true, trim=0.3in 2in 0.4in 1.3in]{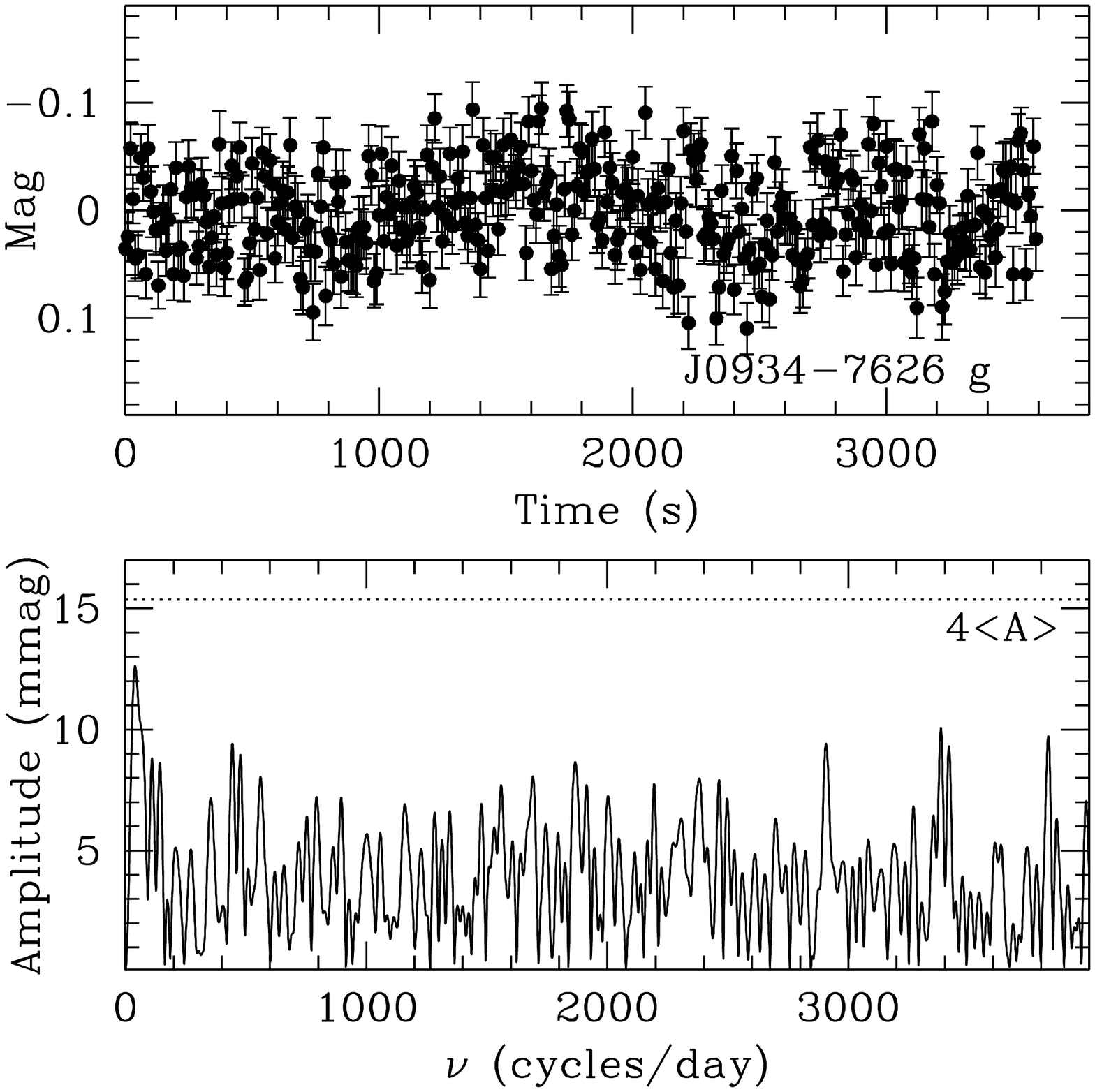}
\includegraphics[width=2.3in, clip=true, trim=0.3in 2in 0.4in 1.3in]{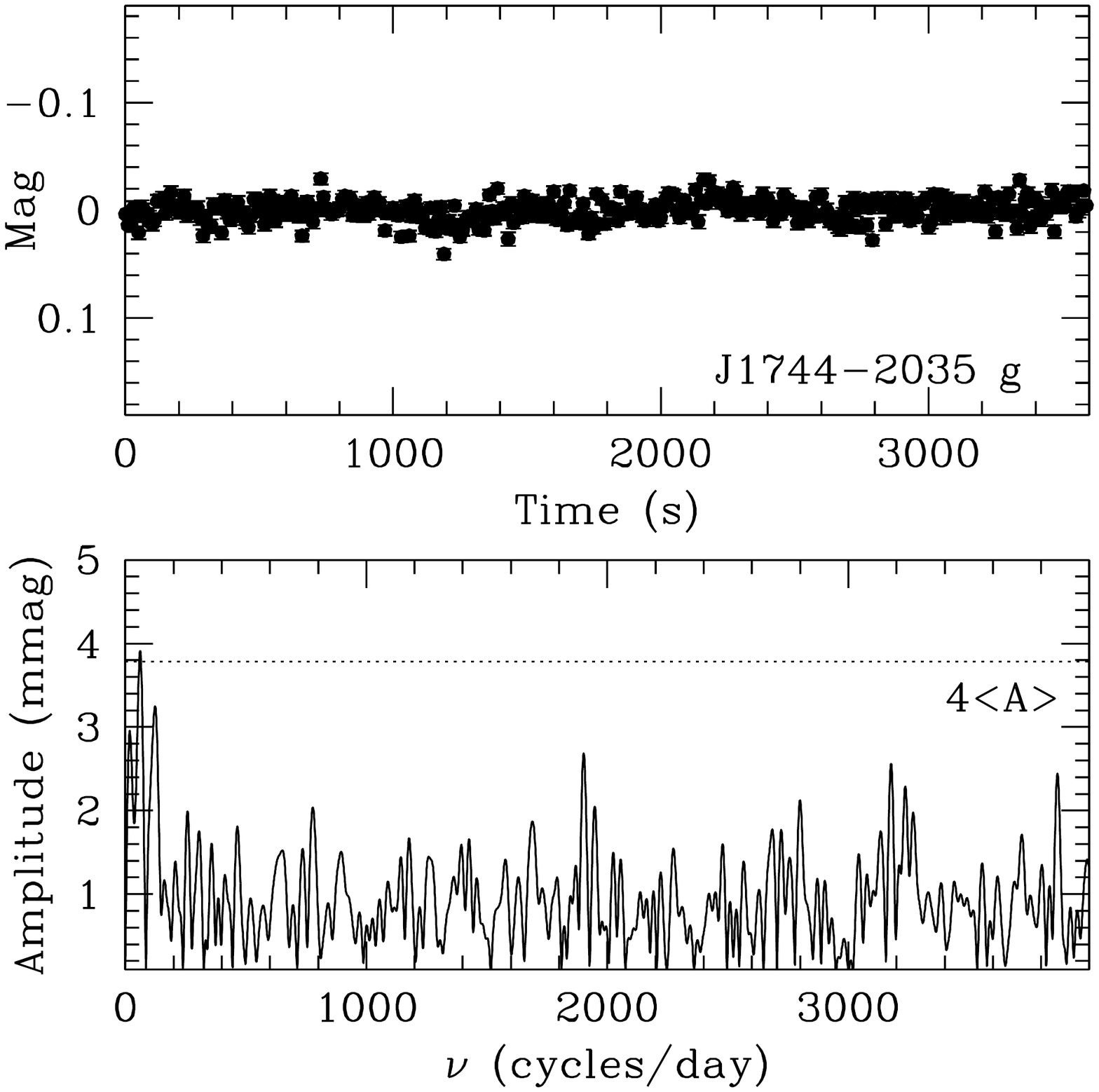}
\caption{Light curves (top) and their Fourier transforms (bottom) of three white dwarfs observed
at the Gemini South telescope with the Zorro instrument. The dotted and dashed lines show
the 4 and 5$\langle {\rm A}\rangle$ level, respectively. The red line in the bottom panel for J0608$-$4515
shows the Fourier transform of one of the reference stars in the Zorro field of view.}
\label{figzorro}
\end{figure*}

Short period photometric variability on minute timescales in single white dwarfs occur due to two main reasons;
pulsation or rapid rotation. Depending on the main atmospheric constituent, white dwarfs pulsate at effective temperatures
near 25,000 K if they have pure helium atmospheres \citep{winget82,vanderbosch22} and 12,000 K if they have pure hydrogen atmospheres
\citep[e.g.,][]{tremblay15}. The most massive DAV pulsators currently known, BPM 37093 \citep{kanaan92} and GD 518 \citep{hermes13} have
$M\approx1.1~M_{\odot}$ \citep{bedard17,kilic20}, and they show multi-periodic oscillations with periods between 400 and 600 s.

Our ultramassive white dwarf sample includes 10 magnetic white dwarfs. \citet{brinkworth13} discovered photometric variability in
67\% of the isolated magnetic white dwarfs in their sample, with periods as short as 27 minutes.  Hence, a significant
fraction of the ultramassive white dwarfs in our sample may show photometric variability due to rapid rotation.
In fact, five of our targets (one DBA and 4 magnetic DAH spectral types) have high-cadence time-series observations available in the literature,
and three show variability at 70 s, 5.88 min, and 6.94 min \citep{kilic21b,pshirkov20,caiazzo21}.
Here we present the results from a search for rapid rotation in the rest of the sample using time-series observations. We discuss
the non-variable objects first, and then present newly discovered rapidly rotating systems, and other potentially variable objects.

\subsection{Nonvariables}

Figure \ref{figapo} shows the APO 3.5m light-curves and their Fourier transforms for 12 targets that were not observed to vary at
minute timescales. The dotted lines mark the 4$\langle {\rm A}\rangle$ level, where $\langle {\rm A}\rangle$
is the average amplitude in the Fourier transform. Depending on the source brightness and the sky conditions, this
limit ranges from 2 millimag in the best case, for J1819$-$1208, to 20 millimag in the worst case, for J0254+3019.
The latter is not ideal, and follow-up observations would be useful to search for low-level variability in J0254+3019 and similar
targets. Our observations typically span two hours, and therefore they do not provide any constraints on the longer
timescale variability of these white dwarfs. Eight of these objects have $g-$band photometry available in the ZTF Data Release 12,
but the ZTF data do not reveal any significant variations either. 

Two of the photometrically non-variable objects shown in Figure \ref{figapo}, J0805$-$1702 and J2255+0710,
are magnetic. \citet{kilic21b} presented
high speed photometry of the latter target over an hour, and ruled out variability at the 16 mmag level. The new
data presented here expand the time baseline to 2 hours and provides more stringent results on the variability
in this system, ruling out variability at 7 mmag and higher.

Figure \ref{figzorro} shows the light curves and their Fourier transforms for three additional targets observed at
the Gemini South telescope with the Zorro instrument. Each object was observed over an hour, and the Zorro
field of view included at least two reference stars that are significantly redder than the target white dwarfs. All
three stars show a peak in the Fourier transform, usually below the 4$\langle {\rm A}\rangle$ level, at low frequencies due to our
observing window and differential extinction. J0608$-$4515 shows the strongest signal at exactly 24 cycles/day and its harmonics. 
The bottom left panel in Figure \ref{figzorro} includes the Fourier transform of one of the reference stars, which
also shows a significant peak at low frequencies, similar to J0608$-$4515. Hence, we classify J0608$-$4515
and the other two objects shown here as non-variable. J0608$-$4515 was also observed as part of the
Catalina Sky Survey. The Catalina data also do not show any large scale variability, though the photometry
is relatively noisy for this star with median errors of 0.2 mag. 

\subsection{Rapid Rotators}

\begin{figure*}
\centering
\includegraphics[width=2.3in, clip=true, trim=0.3in 2in 0.4in 1.3in]{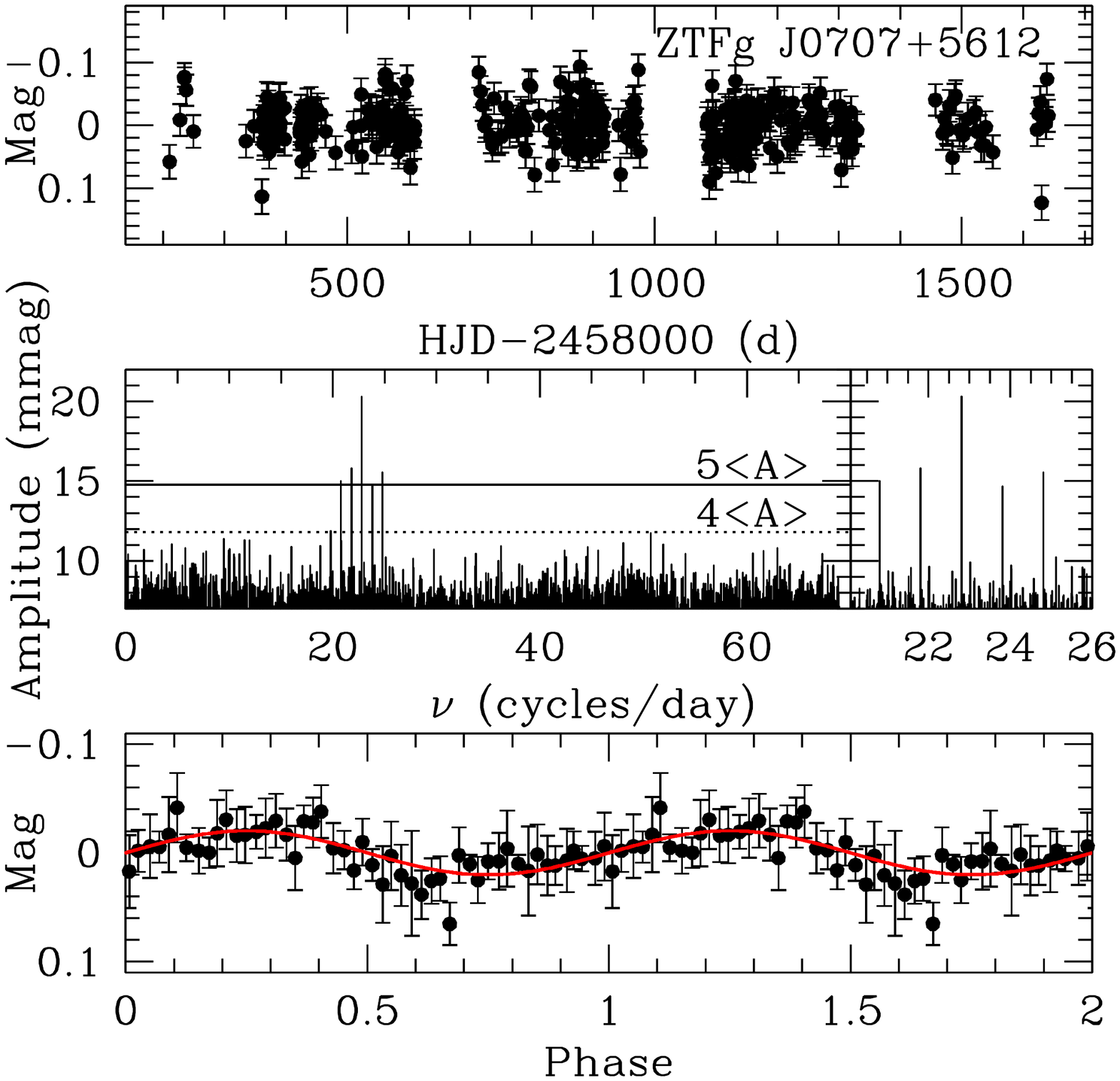}
\includegraphics[width=2.3in, clip=true, trim=0.3in 2in 0.4in 1.3in]{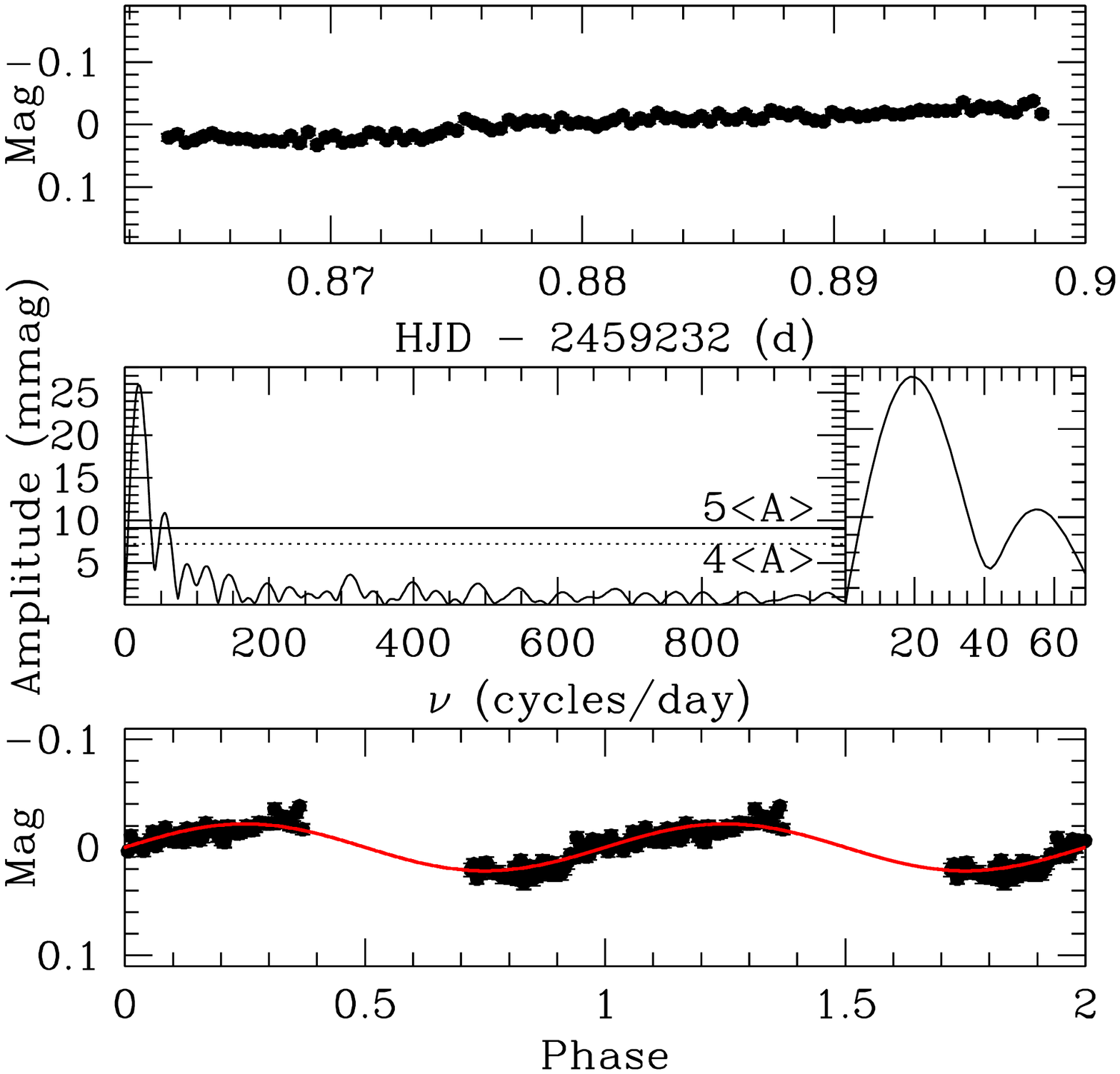}
\includegraphics[width=2.3in, clip=true, trim=0.3in 2in 0.4in 1.3in]{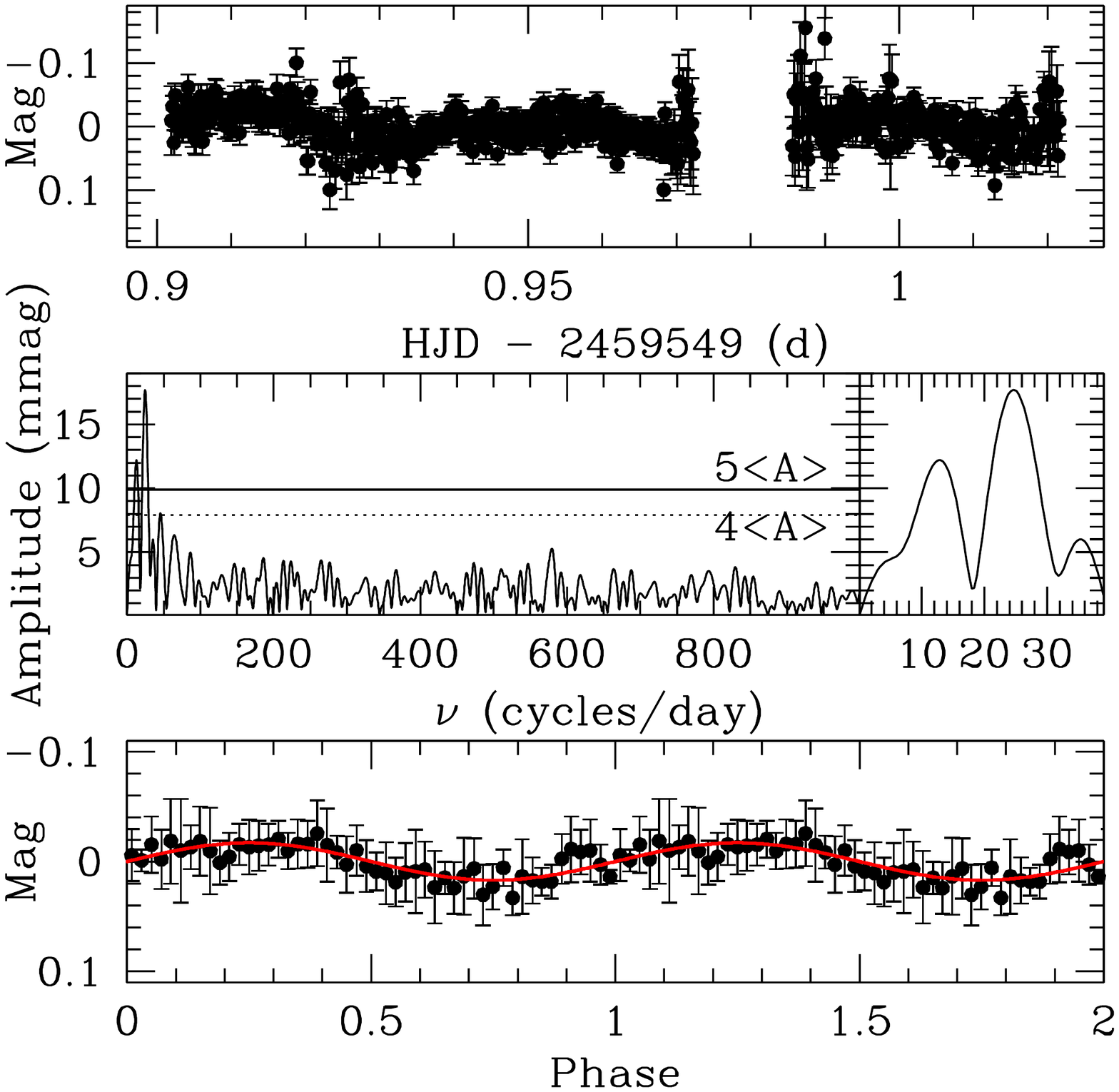}
\caption{ZTF (left), APO (middle), and McDonald (right) light curves of the magnetic white dwarf J0707+5612.
ZTF frequency $22.80963 \pm 0.00004$ cycles/day with $20 \pm 2$ mmag amplitude.
McDonald frequency 24.66 cycles/day with $17 \pm 2$ mmag amplitude, and 
APO frequency 18.7 cycles/day with $22 \pm 1$ mmag amplitude.}
\label{fig0707}
\end{figure*}

We detect evidence of fast rotation in two of our targets, both of which are magnetic.
The first, J0707+5612, is a $T_{\rm eff} = 18100 \pm 350$ K and $M=1.29~M_{\odot}$ (assuming a CO core) white dwarf
with a featureless spectrum that requires strong magnetism to explain the observed spectra.
It shows clear photometric variations in the ZTF data. The left panels in Figure \ref{fig0707} show the
ZTF $g-$band light curve of J0707+5612 along with its Fourier transform, which shows
 a peak with $20 \pm 2$ mmag amplitude at a frequency of $22.80963 \pm 0.00004$ cycles
per day (63 min period). The bottom left panel shows the light curve folded at this highest peak in
the Fourier transform, along with the best-fitting sinusoidal model (red line).

We obtained follow-up BG-40 filter observations of J0707+5612 at both APO and McDonald telescopes.
The middle and right panels in Figure \ref{fig0707} show the results of these observations.
Our APO run was unfortunately limited to an hour, and it is therefore impossible to constrain the
period of variation precisely based on these data. However, the Fourier transform of the APO data
shows a broad peak that is consistent with the ZTF results. Our McDonald 2.1m 
observations span 2.9 hours, and display $17 \pm 2$ mmag amplitude variations at a frequency
of 24.66 cycles per day. Given that the McDonald data cover less than three rotation cycles, the period
estimate is also uncertain. The ZTF data provide the best constraints on the rotation period of J0707+5612.

\begin{figure*}
\centering
\includegraphics[width=3in, clip=true, trim=0.3in 2in 0.4in 1.3in]{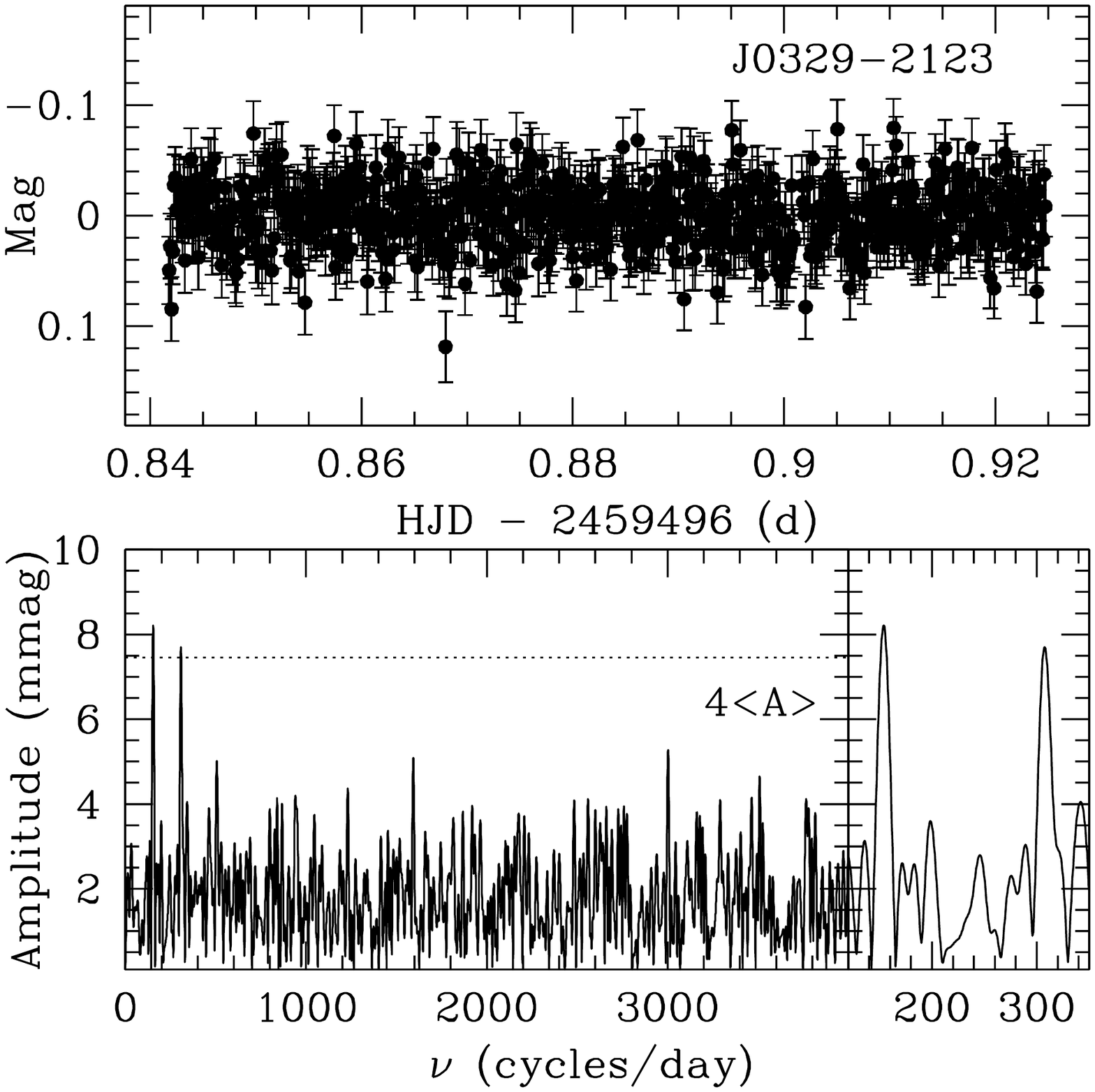}
\includegraphics[width=3in, clip=true, trim=0.3in 2in 0.4in 1.3in]{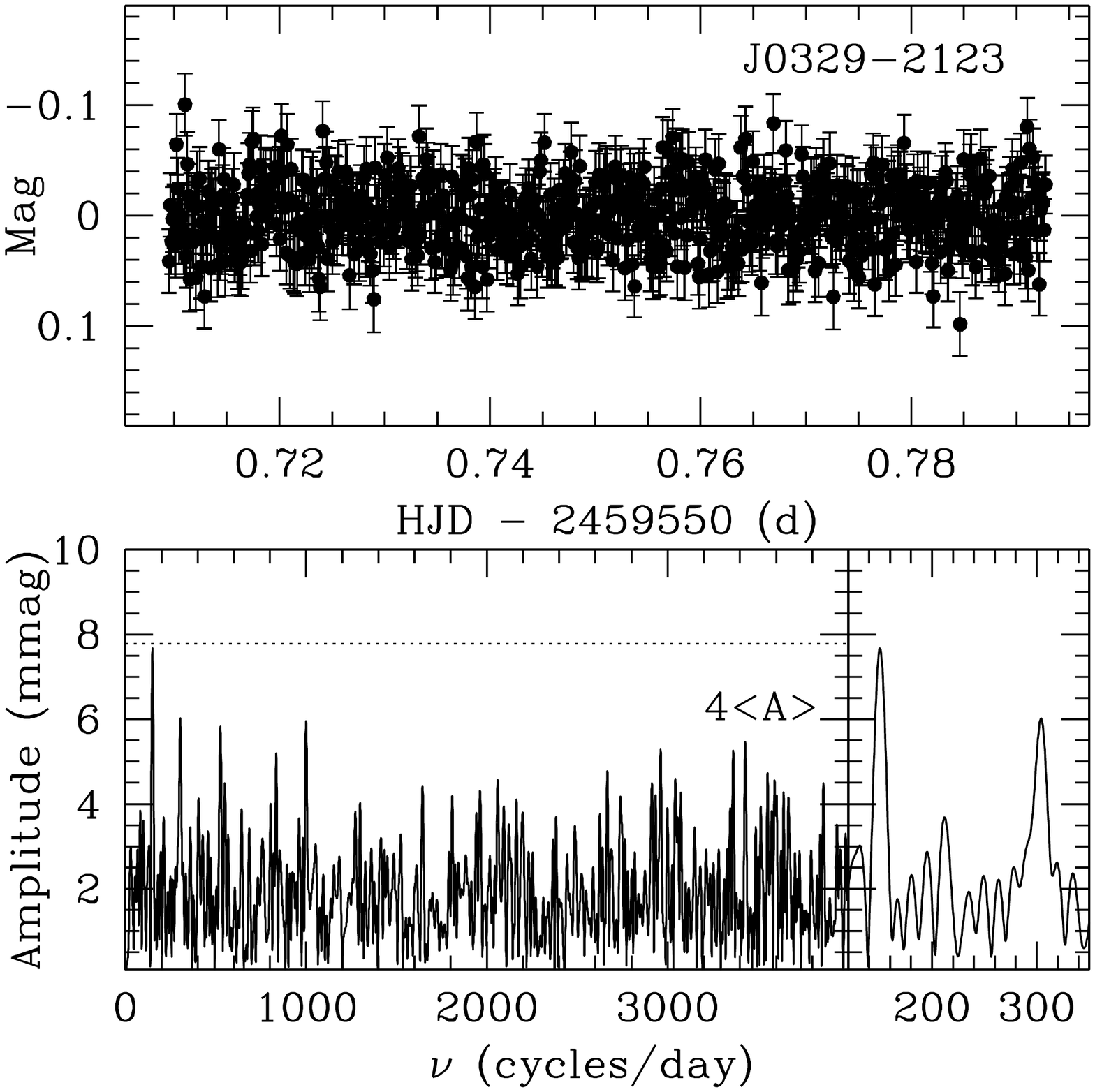}
\caption{APO time-series photometry of the magnetic white dwarf J0329$-$2123 on two separate nights (top panels).
The bottom panels show the Fourier transform of each light curve. The dotted lines show the 4$\langle {\rm A}\rangle$ level.}
\label{fig0329}
\end{figure*}

The second newly discovered rapid rotator is J0329$-$2123, which is also a magnetic DAH white
dwarf with $T_{\rm eff} = 10330 \pm 290$ K and $M=1.34~M_{\odot}$ (assuming a CO core). These parameteres put it outside
of the ZZ Ceti instability strip (see below). Figure \ref{fig0329} shows our APO observations of J0329$-$2123
over two different nights. Each observation is 2 hours long. Observations on UT 2021 Oct 9 (left panels) show
two significant frequencies: the main peak at $154.1 \pm 1.2$ cycles per day (or 9.3 min) with $8.3 \pm 1.4$ mmag amplitude
and its harmonic at $307.3 \pm 1.2$ cycles per day with  $7.8 \pm 1.4$ mmag amplitude. Both of these peaks are
detected at the 4$\langle {\rm A}\rangle$ level. 

The right panels in Figure \ref{fig0329} show the data from UT 2021 Dec 2, with a slightly higher noise level in
the Fourier transform. There is a peak at $150.2 \pm 1.3$ cycles per day with  $7.7 \pm 1.5$ mmag at slightly
below the 4$\langle {\rm A}\rangle$ level, and its first harmonic is detected at $303.8 \pm 1.7$ cycles per day
with $6.1 \pm 1.5$ mmag amplitude. Given the different signal-to-noise ratios of the light curves from each night,
some of these frequencies fall below the 4$\langle {\rm A}\rangle$ level, but they are persistent, and therefore likely to be real. 
Follow-up observations would be useful to confirm the low-level variability seen in this system, and confirm its rotation period of $\sim9.3$ min.

\subsection{Ultramassive ZZ Ceti Candidates}

Our ultramassive white dwarf sample includes several objects near the ZZ Ceti instability strip. Even though the main goal in this
study is not to search for massive pulsating white dwarfs, for completeness we discuss our observations of the DA white dwarfs
near the instability strip. 

\begin{figure}
\center
\includegraphics[width=3.2in, clip=true, trim=0.1in 2in 0.3in 1.3in]{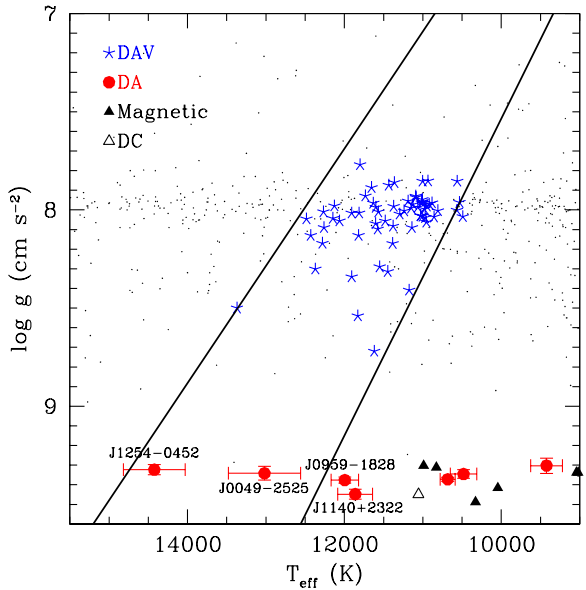}
\caption{Temperatures and surface gravities of the 100 pc sample in the Montreal White Dwarf Database
\citep{dufour17}. Blue stars mark the previously known pulsating DAV white dwarfs, and the solid lines mark
the boundaries of the ZZ Ceti instability strip \citep{tremblay15}. Filled circles, open triangles, and filled triangles
mark the DA, DC, and magnetic white dwarfs in our sample.}
\label{figstrip}
\end{figure}

Figure \ref{figstrip} shows the ZZ Ceti instability strip for DA white dwarfs using the 100 pc MWDD white dwarf sample
\citep{dufour17}. Blue stars mark the previously known pulsating DAV white dwarfs in that sample, and the solid lines
show the empirical boundaries of the instability strip from \citet{tremblay15}. Spectroscopically confirmed DA, DC, and
magnetic white dwarfs in our sample are marked by filled circles, open, and filled triangles, respectively. The DA white
dwarfs near the boundaries of the instability strip are labeled. There are essentially four DA white dwarfs in our sample
that are near the strip: J0049$-$2525, J0959$-$1828, J1140+2322, and J1254$-$0452. 

J0049$-$2525 is by far the best ZZ Ceti candidate in our sample. It was included in our Gemini South observing program, but
unfortunately the high speed photometry component did not get executed in the queue. J1140+2322 and J1254$-$0452 were
observed at APO, and their light curves are included in Figure \ref{figapo}. These stars were observed not to vary down to
approximately 7 and 11 mmag level, respectively. The last object in this
list, J0959$-$1828, is near the red edge of the instability strip. 

\begin{figure*}
\centering
\includegraphics[width=2.3in, clip=true, trim=0.3in 2in 0.4in 1.3in]{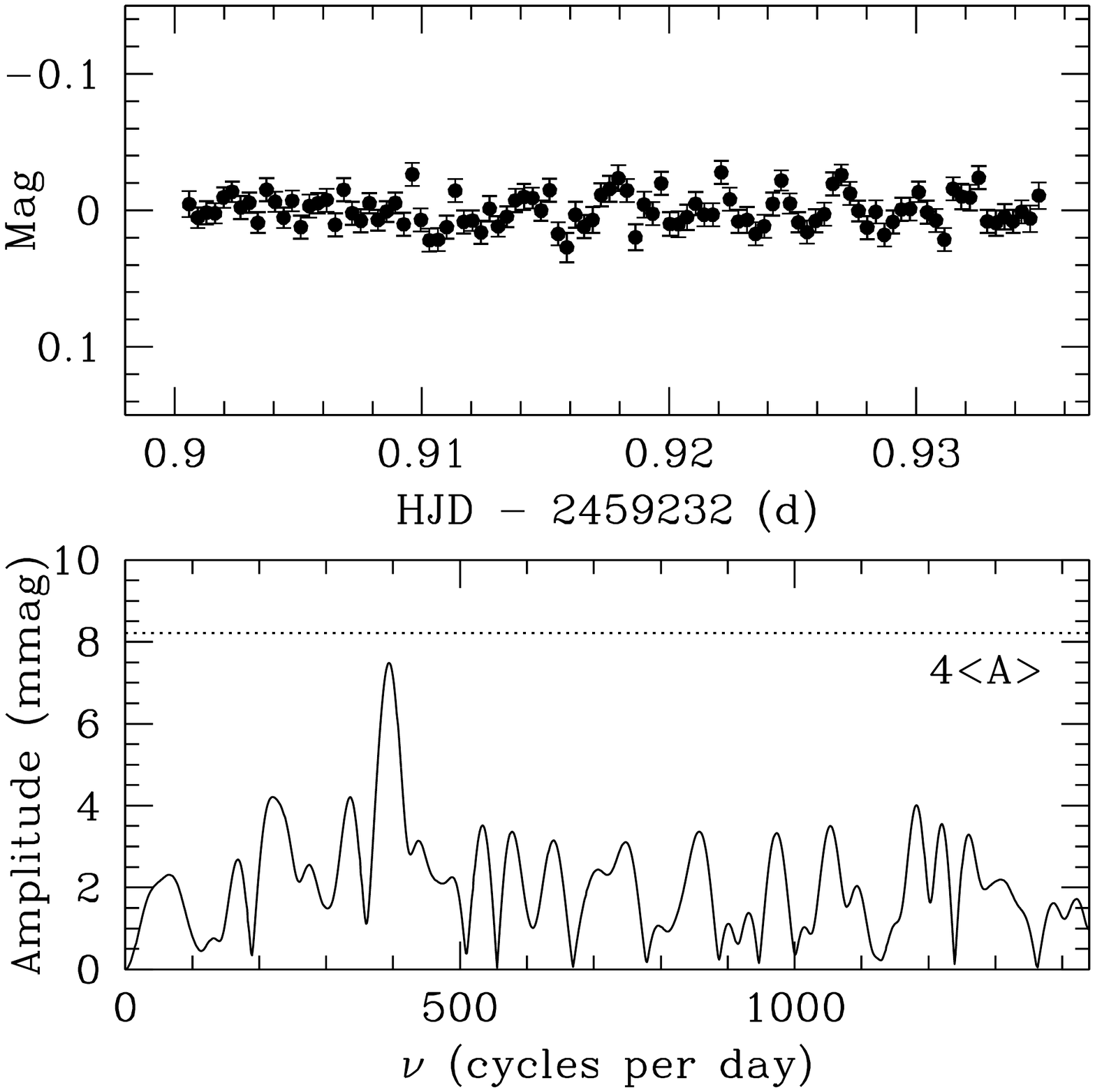}
\includegraphics[width=2.3in, clip=true, trim=0.3in 2in 0.4in 1.3in]{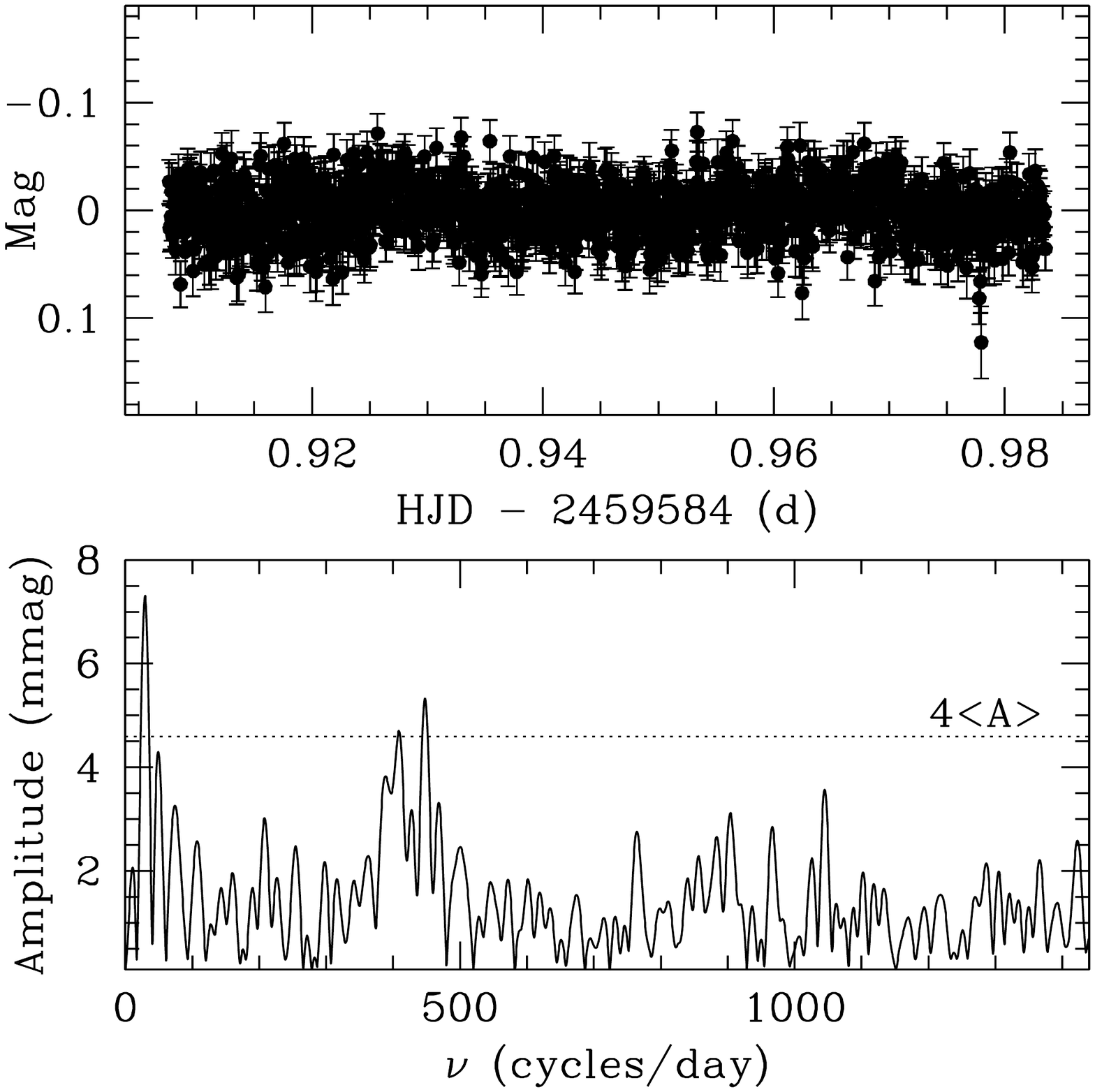}
\includegraphics[width=2.3in, clip=true, trim=0.3in 2in 0.4in 1.3in]{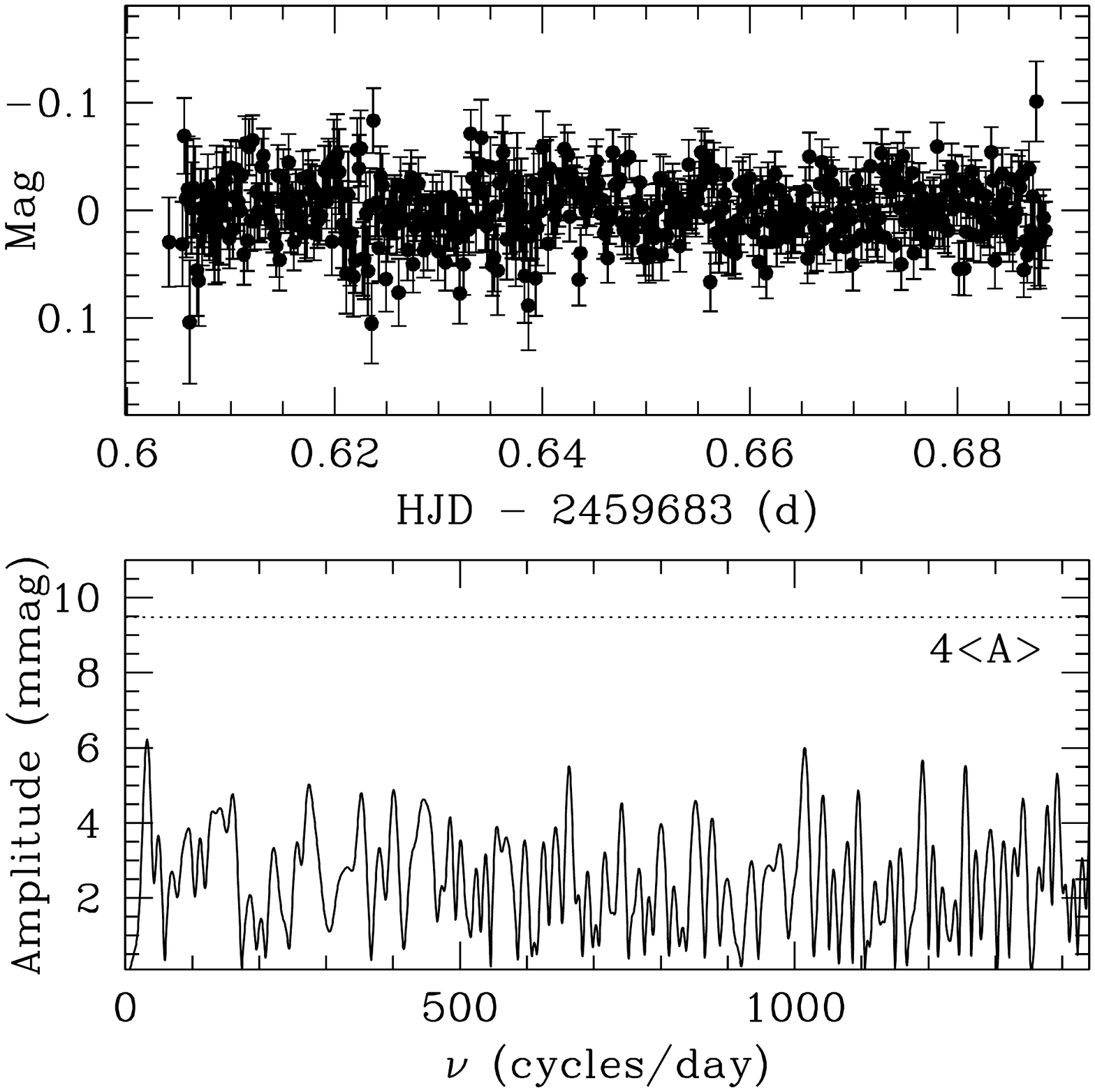}
\includegraphics[width=2.3in, clip=true, trim=0.3in 2in 0.4in 1.3in]{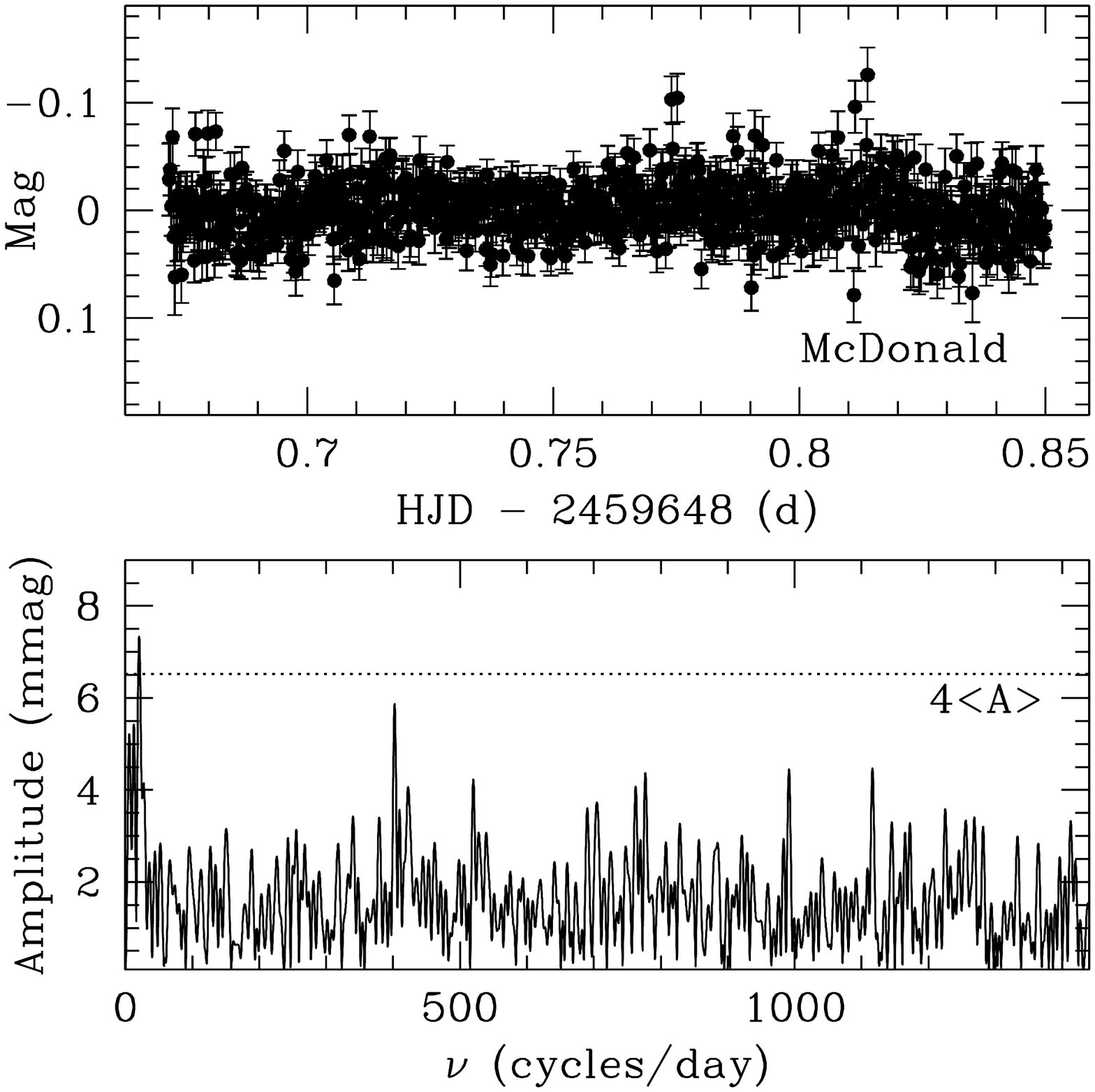}
\includegraphics[width=2.3in, clip=true, trim=0.3in 2in 0.4in 1.3in]{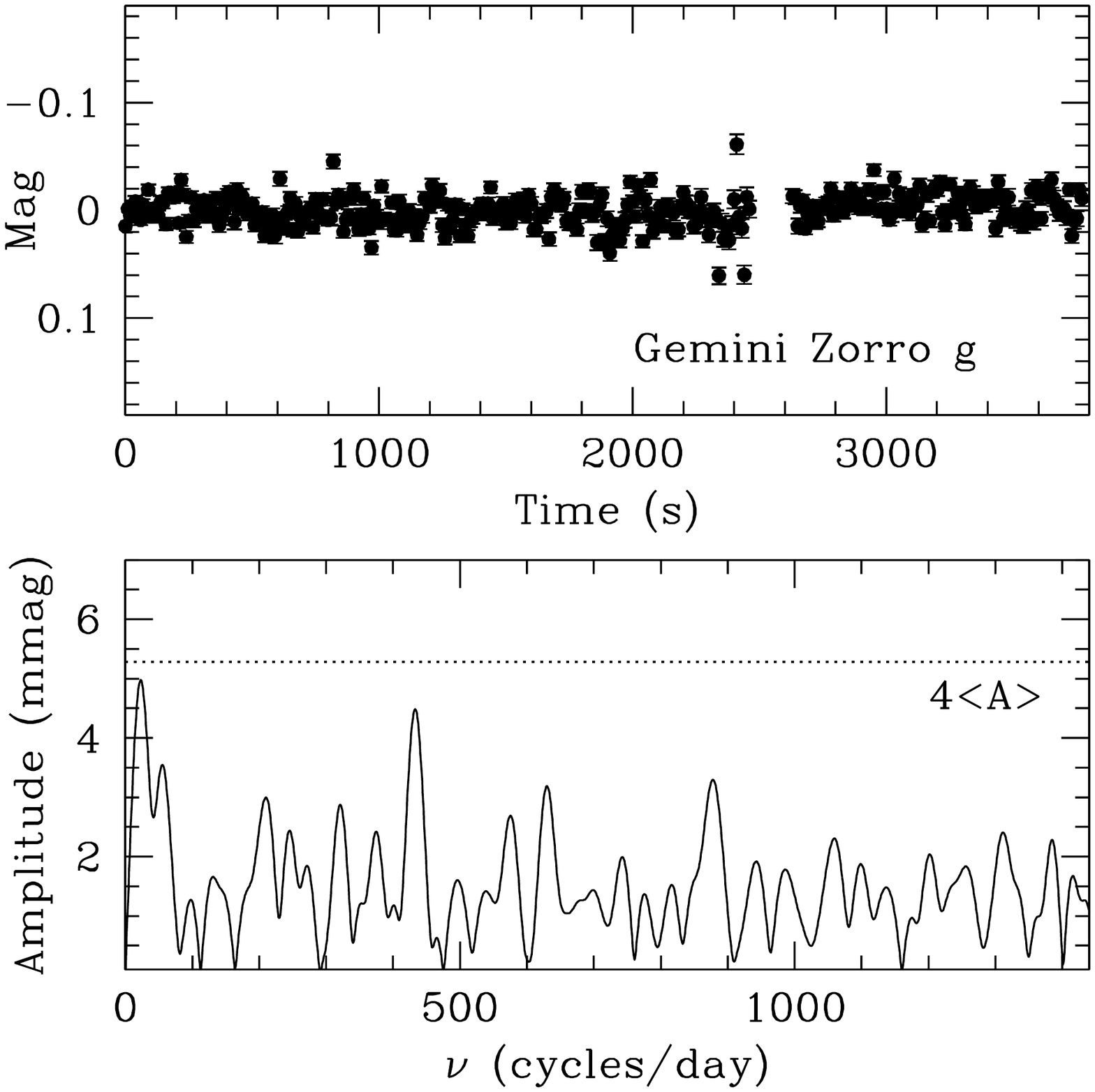}
\caption{APO (top panels), McDonald 2.1m (bottom left panels), and Gemini Zorro (bottom right panels) time-series photometry
of J0959$-$1828 on five different nights. The dotted lines show the 4$\langle {\rm A}\rangle$ level in the Fourier transforms.}
\label{fig0959}
\end{figure*}

We obtained follow-up observations of J0959$-$1828 on five different nights at the APO 3.5m, McDonald 2.1m, and Gemini South
telescopes, with observations spanning one to two hours for APO, one hour for Gemini, and more than four hours for McDonald.
Figure \ref{fig0959} shows all of the light curves for this system. The first night's APO data shows a potential peak near
400 cycles d$^{-1}$ (top left panels). This peak and another near 450 cycles d$^{-1}$ is detected at the 4$\langle {\rm A}\rangle$ level
in the second night's data (top middle panels) as well. However, the data from our third APO night (top right panels) do not show any significant
variability. Longer baseline observations at the McDonald 2.1m also shows a potential peak near 400 cycles d$^{-1}$, but again this
peak is below the 4$\langle {\rm A}\rangle$ level in the Fourier transform. 

Combining all of the APO and McDonald 2.1m, we detect two frequency peaks in the Fourier transform at $402.7 \pm 0.1$ cycles d$^{-1}$
with $4.9 \pm 0.7$ mmag amplitude and $446.4 \pm 0.5$ cycles d$^{-1}$ with $4.4 \pm 0.9$ mmag amplitude. These are detected at
the 4$\langle {\rm A}\rangle$ level in the combined data. On the other hand, our follow-up Gemini Zorro photometry (bottom right panels)
do not show these frequency peaks in the Fourier transform. Hence, we find these data inconclusive in terms of confirming variability in
J0959$-$1828. It is relatively difficult to confirm the potential low-level variability of 4-5 mmag in this system from ground-based observations
at 2-3 m class telescopes. Follow-up time-series photometry on 8m class telescopes or space-based telescopes would be helpful in
confirming any potential variability in J0959$-$1828.

\section{Discussion}

\subsection{The Merger Fraction of Ultramassive White dwarfs}

\begin{figure}
\center
\includegraphics[width=3.4in]{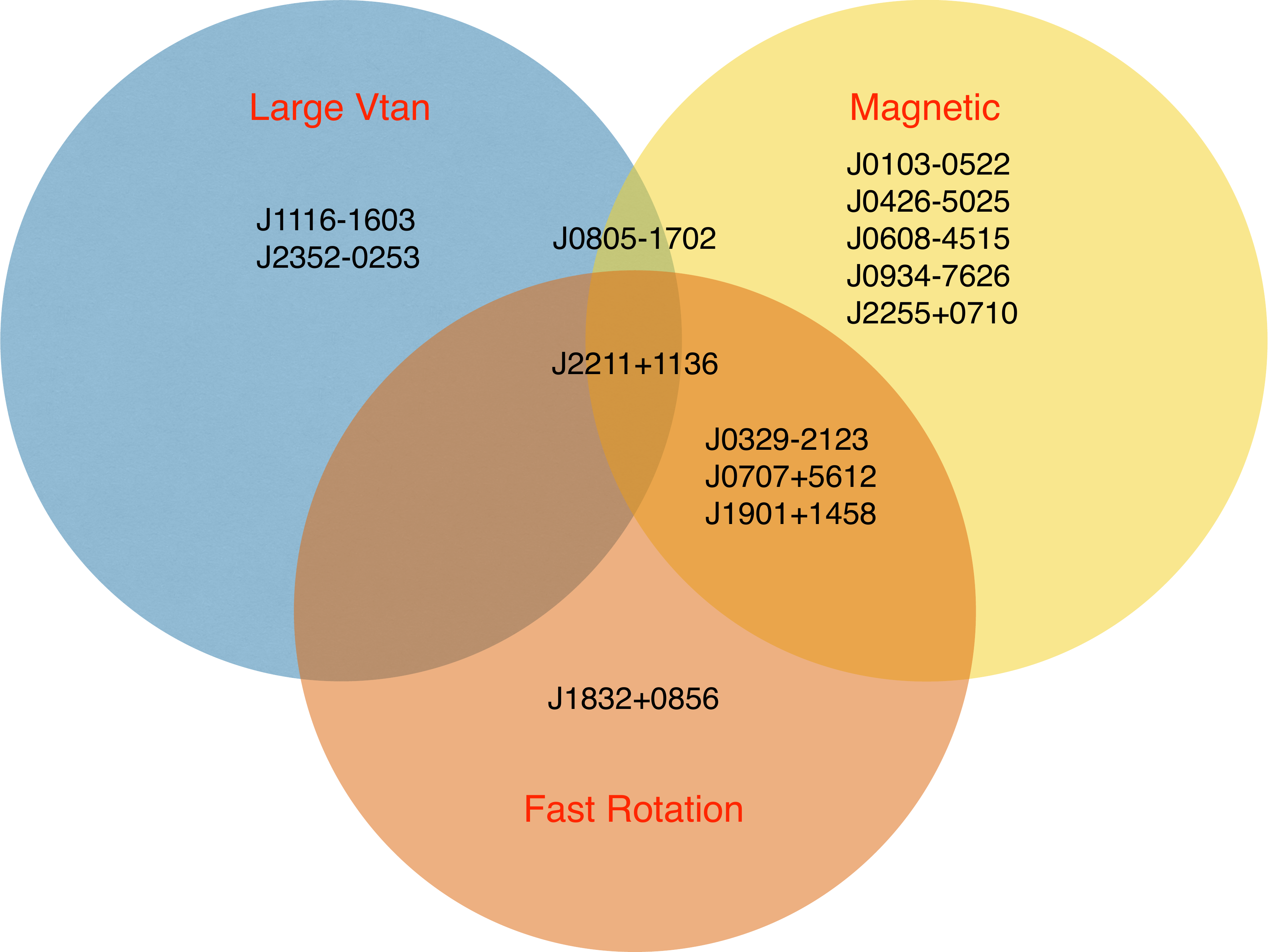}
\caption{Venn diagram comparing the detection of magnetism, rapid rotation, and large tangential velocities in individual objects.
Five targets show more than one symptom of being a merger product, and one object, J2211+1136, shows all three symptoms.}
\label{figvenn}
\end{figure}

We searched for merger products among the 25 ultramassive white dwarfs with $M\geq1.3~M_{\odot}$ identified by \citet{kilic21}.
We used three main methods for this; we searched for evidence of magnetism through optical spectroscopy, evidence of rapid
rotation through high cadence photometry, and evidence of unusual kinematics through Gaia astrometry. Figure \ref{figvenn} shows
a Venn diagram comparing the detection of magnetism, rapid rotation, and large tangential velocities in individual objects.

Our follow-up spectroscopy shows that 10 of these 25 objects are
strongly magnetic, with field strengths of 5 MG or higher. This fraction, 40\%, is a factor of four higher than observed in the local
20-25 pc white dwarf sample \citep{holberg16,hollands18}, and is telling about the origin of these ultramassive white dwarfs.

\citet{isern17} discussed the origin of magnetism in white dwarfs and proposed that in addition to the commonly invoked fossil fields
and binary interactions, magnetic fields can also arise during core crystallization \citep[see the review by][]{ferrario20}. In this scenario,
the crystallization of a C/O core
white dwarf and the ensuing phase separation leads to the formation of a convective mantle on top of a solid core. This convective
region can produce a dynamo with magnetic field strengths of up to 0.1 MG. All three scenarios may be needed to explain the
frequency of magnetic white dwarfs in volume limited white dwarf samples. However, the fossil fields and the crystallization scenarios
are clearly inadequate for explaining the rapidly rotating and strongly magnetic ($B\geq5$ MG) ultramassive white dwarfs in our sample.
\citet{garciaberro12} demonstrated that the hot, convective, and differentially rotating corona present in the outer layers of a merger
remnant can produce strong magnetic fields \citep[see also][]{tout08,briggs15}, and such objects are expected to rotate on $\sim$minute
timescales \citep{schwab21}. 

Our high cadence observations at APO, McDonald, and Gemini identified two additional rapidly rotating white dwarfs that are also
strongly magnetic. J0707+5612 and J0329$-$2123 show photometric variability with periods of 63 min and $\approx9.3$ min, respectively.
These are much faster than the typical day long rotation rates observed in average mass ($M\sim0.6~M_{\odot}$) pulsating white dwarfs
\citep{hermes17a}, but they are consistent with the expectations for white dwarf merger products \citep{schwab21}. 
There are three other rapidly rotating white dwarfs known in our sample, J1832+0856, J1901+1458, and J2211+1136
\citep{pshirkov20,caiazzo21,kilic21b}, bringing the total number of rapidly rotating objects in our sample to five.
Hence, 20\% of the most massive white dwarfs in the Montreal White Dwarf Database 100 pc sample rotate with periods ranging from
roughly 1 min to 1 hour. 

Kinematics provide another way to identify white dwarfs with unusual evolutionary histories. Binary interactions and mergers
can reset the evolutionary clock of a white dwarf progenitor and make its descendant white dwarf appear younger and hotter than what it would be if it
went through single star evolution. Our ultramassive white dwarf sample consists of objects with relatively young cooling ages
of $\sim1$ Gyr. Hence, they should, on average, show disk kinematics. \citet{kilic21} identified four outliers in tangential velocity.
Even though the average tangential velocity of the sample, 21 km s$^{-1}$, is consistent with a young disk population,
they found four objects with $V_{\rm tan}>50$ km s$^{-1}$. These four objects, J0805$-$1702, J1116$-$1603,
J2211+1136, and J2352$-$0253, likely suffered from binary interactions and mergers in the past. 
 
Five objects show more than one symptom of being a merger product. Out of the 10 magnetic white dwarfs in our sample,
four are also fast rotators, and two display large tangential velocities. Namely, J0329$-$2123, J0707+5612, and J1901+1458
are rapidly rotating, magnetic ultramassive massive white dwarfs. J0805$-$1702 is a magnetic ultramassive white dwarf with
an unusually large tangential velocity of 90 km s$^{-1}$. J2211+1136 is the best example of a merger product. It is a $1.27~M_{\odot}$
white dwarf with a $B=15$ MG field, a rotation period of only 70 s, and a relatively large tangential velocity of 56 km s$^{-1}$. 

The Venn diagram in Figure \ref{figvenn} excludes the normal DA and DC white dwarfs in our sample. However, there is one more object
that is likely a merger product, but it is excluded from this figure because it does not show any obvious evidence of magnetism,
rapid rotation, or a large tangential velocity. The hot DQ white dwarf J1819$-$1208 is likely a merger product based on its unusual
composition and the overall properties of the hot DQ population in the solar neighborhood. More than 70\% of the hot DQs are
magnetic, and at least one third of these stars are also variable \citep[e.g.,][]{montgomery08,dufour11}.
\citet{dunlap15} argued that the unique atmospheric compositions, high masses, high incidence of magnetism, and relatively high
tangential velocities favor a merger origin for hot DQ white dwarfs. \citet{williams16} demonstrate that the photometric variability in
these objects is likely due to rotation, and that hot DQ white dwarfs contain many rapid rotators. Our time-series photometry on
J1819$-$1208 did not reveal short period variability, but our observations are not sensitive to hour or day long periods.
Hence, the unusual atmospheric composition of J1819$-$1208 makes it a prime candidate for a merger product.

Adding J1819$-$1208 to the list of 13 objects shown in Figure \ref{figvenn} brings the total number of merger products
among our ultramassive white dwarf sample to 14, which is remarkable. We use the binomial probability distribution to
compute the upper and lower limits on the frequency of mergers \citep{burgasser03}. Since this probability
function is not symmetric about its maximum value, we report the range in probability that delimits 68\% of the integrated
probability function, equivalent to $1\sigma$ Gaussian limits. Since 14 of the 25 of our targets show evidence of a merger origin,
this corresponds to a merger fraction of $56^{+9}_{-10}$\%, and a two sigma lower limit of 36.6\%.

\subsection{Binary Population Synthesis Predictions}

From binary population synthesis calculations, 
\citet{temmink20} estimated that 10 to 30\% of all observable single white dwarfs are formed through binary mergers, where
the primary contribution is from the mergers between a post-main-sequence star and a main-sequence star \footnote{These mergers
contribute 45\% of the systems in their default model.}. Mergers provide a more significant contribution
to single massive white dwarfs. For masses above $0.9~M_{\odot}$ mergers contribute 30 to 45\% of all observable single white
dwarfs within 100 pc, where the dominant contribution is from double white dwarf mergers in most models.

The outcome of the binary population synthesis calculations depends heavily on the input assumptions for the binary physics and
initial conditions (e.g. the ranges in the fractions mentioned above). One of the most important assumptions for the formation of
compact binaries and mergers involve the common envelope phase. The $\alpha$
prescription \citep{webbink84} is most commonly used to model it, where $\alpha$ represents the fraction of the orbital energy
that is used to unbind the common envelope. An additional dimensionless parameter $\lambda$ \citep{dekool87}, which depends on the structure
of the donor star, is also used in this prescription \citep{nelemans00,temmink20}. $\alpha$ and $\lambda$ are often treated
as a single parameter $\alpha \lambda$ since the product of the two parameters, $\alpha \lambda$, appears naturally in solutions
for the orbital separation of the binary system after the common envelope evolution. 
In the fiducial model of \citet{temmink20}, they adopt $\alpha \lambda = 2$ based on the reconstruction of the evolution of double helium white dwarfs
by \citet{nelemans00}. They also tested the sensitivity of their models to a less efficient and a more efficient common envelope phase
by assuming $\alpha \lambda = 0.25$ and $\alpha \lambda = 5$, respectively\footnote{Even though the common envelope efficiency parameter
$\alpha$ is not well known, the structure parameter $\lambda$ is constrained better \citep[see the discussion in][]{toonen13}. Hence,
a small value of  $\alpha \lambda$ corresponds to a small value of $\alpha$, or an inefficient common envelope.}. 
 
Figure 9 in \citet{temmink20} shows the merger fraction as a function of mass based on binary population synthesis calculations using
six different prescriptions (the default model, efficient $\alpha$, and inefficient $\alpha$, etc). The largest difference in the predicted merger fraction is indeed due to the adopted common envelope prescription
For example, the default model predicts a 37\% merger rate for
$1.3~M_{\odot}$ white dwarfs, but this rate goes down to about 28\% for an efficient common envelope with $\alpha \lambda = 5$, and it
could be as high as 44\% for an inefficient common envelope with $\alpha \lambda = 0.25$. 

\subsection{Implications for the Common-Envelope Evolution}

The merger fraction of our ultramassive white dwarf sample with $M\approx1.3~M_{\odot}$ is $56^{+9}_{-10}$\%. This is higher than all of the
predictions from the binary population synthesis calculations of \citet{temmink20}, but it is consistent with the models using an inefficient (low)
common envelope parameter within $\approx 1\sigma$. The default model in the population synthesis calculations is at the $2\sigma$ lower
limit of our measurement, and the calculations assuming an efficient common envelope evolution are at the $3\sigma$ lower limit of our
measurement. Hence, the observed merger fraction of our ultramassive white dwarf sample clearly favors low values of the common envelope
efficiency.

There is additional support for low values of the common envelope efficiency from other systems studied in the literature. 
Reconstructing the evolution of post-common-envelope binaries of white dwarfs with main-sequence star companions,
\citet{zorotovic10} found that most systems can be explained by a broad range of $\alpha$ values, but they found simultaneous
solutions for all post common envelope binaries in their sample only for $\alpha=$ 0.2-0.3 \citep[also see][]{camacho14}.
\citet{toonen13} compared the synthetic and observed population of visible post common envelope binaries in the SDSS
and also concluded that common envelope efficiency parameter must be low. Finally, \citet{zorotovic22} reconstructed
the evolutionary histories of post common envelope binaries with brown dwarf companions, and concluded that
the vast majority of post common envelope binaries can be described with a small efficiency parameter.

Constraining the value of $\alpha$ has significant implications for understanding the white dwarf merger rate and the outcome
of the common envelope evolution. For example, \citet{temmink20} calculated an integrated merger rate (that leads to observable
single white dwarfs) ranging from 0.013 to 0.032 $M_{\odot}^{-1}$. The upper bound here is for a small common envelope efficiency
parameter. The corresponding Galactic rate is 0.04 to 0.09 per year. The relatively high merger fraction that we found in our
ultramassive white dwarf sample favors merger rates closer to the upper limit of this estimate.

\section{Conclusions}

We present the results from a comprehensive spectroscopic and photometric survey of the 25 ultramassive white dwarfs with
$M\approx1.3~M_{\odot}$ identified by \citet{kilic21}. We use rapid rotation, kinematics, magnetism, and unusual atmospheric
composition to identify merger candidates. We found 10 magnetic white dwarfs with field strengths ranging from a few MG to hundreds
of MG, four of which have rotation periods in the minute to hour range. Four systems show large tangential velocities, and one object
is a hot DQ white dwarf with a carbon dominated atmosphere. Several of our targets show multiple symptoms of being a merger product.
J2211+1136 is the best example, it is ultramassive, highly magnetic, it rapidly rotates, and has a relatively large tangential velocity. 

In total, we identify 14 objects out of 25 as likely merger systems, which implies a merger fraction of $56^{+9}_{-10}$\%. This fraction is
higher than the predictions from the binary population synthesis calculations, but is closest to the models assuming a low
common envelope efficiency parameter $\alpha$. Hence, our results provide further support to the low common envelope
efficiency suggested by other authors studying post common envelope binaries of white dwarf plus main-sequence or brown
dwarf companions. 

Our follow-up photometric survey was designed to be inclusive of all targets, including normal DA stars, so that we do not miss
any rapidly rotating systems. However, excluding the DA white dwarfs near the ZZ Ceti instability strip, we detected
short period variability only among the magnetic white dwarfs in the sample. There is one more rapidly rotating system,
J1832+0856 \citep{pshirkov20}, which is a DBA white dwarf. Hence, all of the rapidly rotating systems in our sample are either magnetic
or non-DA white dwarfs. Phase-resolved spectroscopy of these rapidly rotating systems would be helpful for understanding the source
of variability, whether it is due to a chemically inhomogeneous surface composition, rotational modulation of a complex magnetic field,
and/or spots \citep[e.g.,][]{dupuis00,kilic19b,caiazzo21}. 

Even though our survey of 25 stars provides the first reliable observational constraints on the merger fraction of single ultramassive
white dwarfs, a larger spectroscopic survey will be essential for increasing the sample size and extending the mass range probed. 
The Gaia EDR3 white dwarf catalog \citep{gentile21} includes 34 candidates with parallax $\varpi\geq10$ mas, $M\geq1.3~M_{\odot}$,
and $T_{\rm eff}\geq8000$ K based on the pure hydrogen atmosphere model fits to the Gaia photometry and parallax.  This number
goes up to 324 candidates if we remove the parallax constraint. Follow up observations of such a sample can provide more precise
estimates of the merger fraction of ultramassive white dwarfs.  

Vera Rubin Observatory's Legacy Survey of Space and Time (LSST) will provide an unprecedented opportunity to identify
the merger products among the solar neighborhood white dwarfs. The LSST will deliver $\sim$nightly cadence photometry
for millions of white dwarfs and also parallaxes and proper motions for faint but nearby objects. Hence, the LSST will
find both rapidly rotating white dwarfs and faint white dwarfs with large tangential velocities. The LSST, along
with the upcoming large scale spectroscopic surveys like the Dark Energy Spectroscopic Instrument (DESI)
Milky Way Survey \citep{allende20} and the SDSS-V \citep{kollmeier19}, will significantly improve the merger fraction
constraints, with implications for further constraining the physics of the common envelope evolution.
 
\section*{Acknowledgements}

This work is supported in part by the NSF under grants  AST-1906379 and AST-2205736, the NASA under grant 80NSSC22K0479, the NSERC Canada,
the Fund FRQ-NT (Qu\'ebec), and by the Smithsonian Institution. 
AK acknowledges support from NASA through grant 80NSSC22K0338.

The Apache Point Observatory 3.5-meter telescope is owned and operated by the Astrophysical Research Consortium.

This paper includes data gathered with the 6.5 meter Magellan Telescopes located at Las Campanas Observatory, Chile.

Based on observations obtained at the international Gemini Observatory, a program of NSF's NOIRLab, which is managed by the Association of Universities for Research in Astronomy (AURA) under a cooperative agreement with the National Science Foundation on behalf of the Gemini Observatory partnership: the National Science Foundation (United States), National Research Council (Canada), Agencia Nacional de Investigaci\'{o}n y Desarrollo (Chile), Ministerio de Ciencia, Tecnolog\'{i}a e Innovaci\'{o}n (Argentina), Minist\'{e}rio da Ci\^{e}ncia, Tecnologia, Inova\c{c}\~{o}es e Comunica\c{c}\~{o}es (Brazil), and Korea Astronomy and Space Science Institute (Republic of Korea).

\section*{Data availability}

The data underlying this article are available in the MWDD at
http://www.montrealwhitedwarfdatabase.org and in the Gemini Observatory Archive at
https://archive.gemini.edu, and can be accessed with the program numbers
GN-2022A-Q-303, GS-2022A-Q-106, and GS-2022A-Q-303. 
The APO and McDonald Observatory data that support the findings of this study are available
from the corresponding author upon reasonable request.

\input{ms.bbl}

\bsp
\label{lastpage}

\end{document}